\documentclass[conference]{IEEEtran}
\usepackage[utf8]{inputenc}
\IEEEoverridecommandlockouts
\usepackage{cite}
\usepackage{amsmath,amssymb,amsfonts}
\usepackage{algorithmic}
\usepackage{graphicx}
\usepackage{float}
\usepackage{epstopdf}
\usepackage{subfigure} 
\usepackage{textcomp}
\usepackage{xcolor}
\usepackage{csvsimple}
\def\BibTeX{{\rm B\kern-.05em{\sc i\kern-.025em b}\kern-.08em
    T\kern-.1667em\lower.7ex\hbox{E}\kern-.125emX}}

\begin{document}

\title{Evaluation of Surrogate Models for Multi-fin Flapping Propulsion Systems}

\author{
  \IEEEauthorblockN{
    Kamal Viswanath\IEEEauthorrefmark{1},
    Alisha Sharma\IEEEauthorrefmark{1},
    Saketh Gabbita\IEEEauthorrefmark{2},
    Jason Geder\IEEEauthorrefmark{1},
    Ravi Ramamurti\IEEEauthorrefmark{1}, and
    Marius Pruessner\IEEEauthorrefmark{3}%
    \thanks{\IEEEauthorrefmark{2}Science and Engineering Apprenticeship Program (SEAP) student}%
    \thanks{DISTRIBUTION A:  Approved for public release, distribution is unlimited.}
  }
  \IEEEauthorblockA{
    \IEEEauthorrefmark{1}
    \IEEEauthorrefmark{2}
    \textit{Laboratories for Computational Physics and Fluid Dynamics},\\
    Naval Research Laboratory, Washington, DC. 20375
  }
  \IEEEauthorblockA{
    \IEEEauthorrefmark{3}
    \textit{Center for Bio-molecular Science and Engineering},\\
    Naval Research Laboratory, Washington, DC. 20375
  }
}

\maketitle


\begin{abstract}
The aim of this study is to develop surrogate models for quick, accurate prediction of thrust forces generated through flapping fin propulsion for given operating conditions and fin geometries. Different network architectures and configurations are explored to model the training data separately for the lead fin and rear fin of a tandem fin setup. We progressively improve the data representation of the input parameter space for model predictions. The models are tested on three unseen fin geometries and the predictions validated with computational fluid dynamics (CFD) data. Finally, the orders of magnitude gains in computational performance of these surrogate models, compared to experimental and CFD runs, vs their tradeoff with accuracy is discussed within the context of this tandem fin configuration.
\end{abstract}



\section{Introduction}

To address the need for more effective and efficient maneuvering in marine environments, propulsion and control systems inspired by fish and other aquatic organisms are starting to provide viable alternatives to traditional vehicle thrusters and control surfaces in a range of underwater regimes.  Research to identify the principles of fish locomotion and characterize the propulsive performance for a variety of artificial, bio-inspired underwater propulsion systems has steadily grown over the past few decades with the predominant focus on flapping fins or foils, in various configurations, to achieve thrust.  However, while biologists and engineers have worked together to study and take inspiration from nature in the development of robotic fins, research seeking to understand and model the propulsive performance of multiple fins operating on a vehicle is limited.

Biologists have observed the coordinated body and fin motions exhibited by various fish species and studied the wake interactions between these moving surfaces \cite{hove_boxfishes_2001,lauder_learning_2006,flammang_volumetric_2011}.  Various research groups have studied a multitude of fin shapes, stroke parameters, and configurations, as well as different fin materials and surface curvature control techniques, for robotic underwater propulsion systems \cite{barrett_drag_1999,licht_design_2004,kato_design_2008,sitorus_design_2009,kato_elastic_2008,palmisano_robotic_2012,esposito_robotic_2012,moored_investigating_2008}, and some have derived reduced order models for individual fins to capture relevant features at significantly lower computational cost \cite{bozkurttas_understanding_2008}.  While there is a rich set of parametric studies of isolated flapping fins, research on the propulsive performance of a system of multiple fins is limited including studies of interactions between dorsal and caudal fins \cite{akhtar_hydrodynamics_2007,mignano_passing_2019}, tandem pectoral fins \cite{ramamurti_computational_2018,geder_underwater_2017}, and pectoral and caudal fins \cite{ramamurti_propulsion_2019}.  Reduced order models of these multi-fin systems is even less common including quasi-steady models of tandem flapping foils \cite{geder_four-fin_2011,muscutt_performance_2017}.  To create an accurate and time-effective tool for the design, analysis and control of vehicles propelled by artificial flapping fins, comprehensive studies that identify the effects of various fin parameters on multi-fin flow interactions must be used to develop lower computational cost surrogate models of forces and propulsive efficiencies.

To characterize the effects of flow interactions between oscillating propulsive surfaces caused by time-varying wake structures, we have studied a configuration of tandem, identical geometry fins flapping perpendicular to the direction of flow \cite{ramamurti_computational_2018,geder_underwater_2017}, similar to lift-based pectoral fin motions of some fish species \cite{liebe_diversity_2006}.  Using the results of these previous studies, the current research seeks to design and evaluate surrogate models of stroke-averaged and time-varying thrust for this multi-fin system. The goal is to develop models that could give a quick, accurate force profile given a fin configuration and a set of kinematics, without explicitly embedding the mathematics within it. Beyond accuracy, we seek models that are data efficient to train and generalizable, so we can be confident in tests of different fin configurations.

\begin{figure}[htbp]
\centerline{\includegraphics[width=3.5in,keepaspectratio]{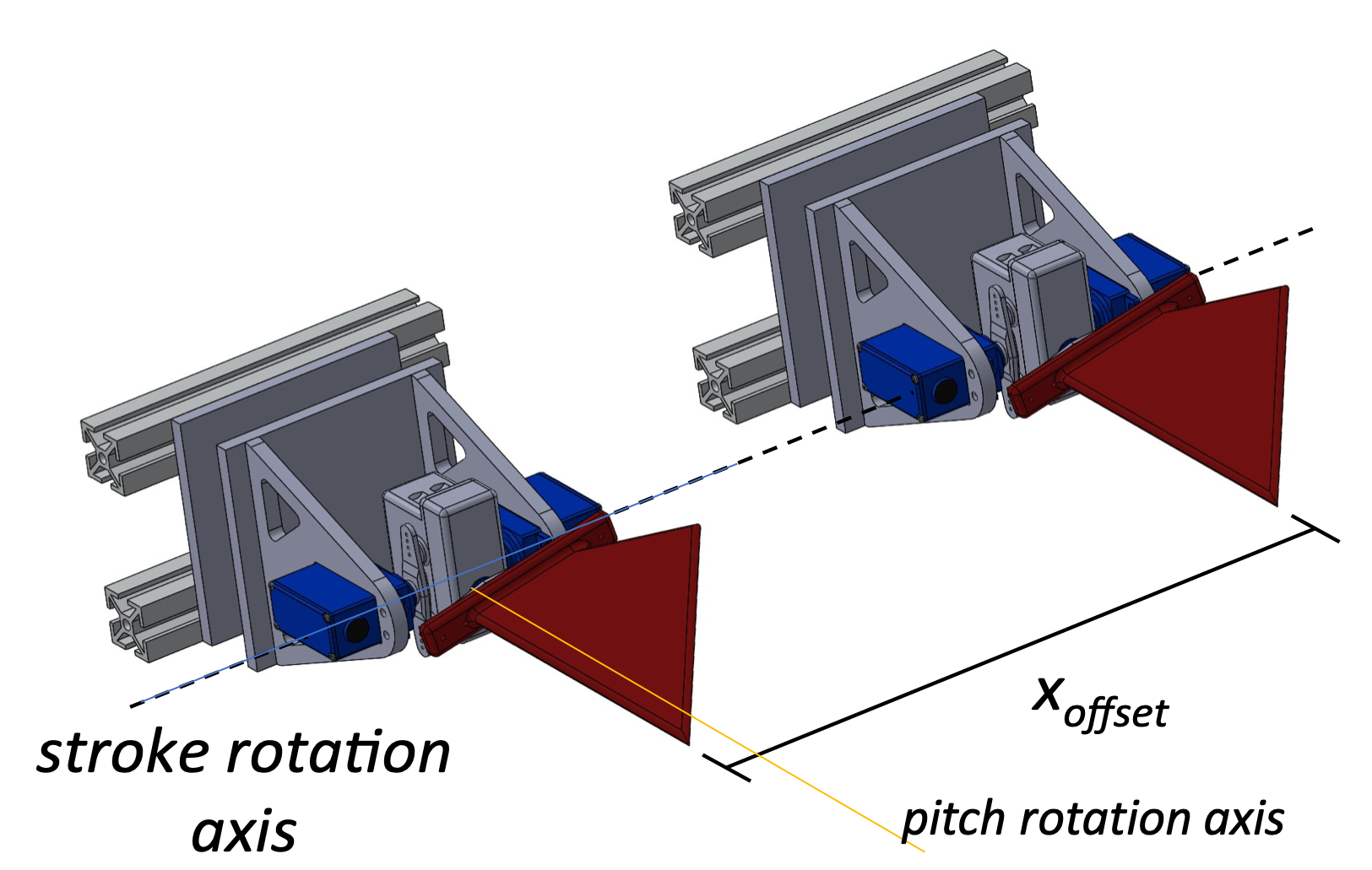}}
\caption{Tandem fin configuration.}
\label{fig-tandemfin}
\end{figure}

As shown in fig.~\ref{fig-tandemfin}, the propulsion system consists of two fins in tandem and on an underwater test vehicle there will be one set of the above configuration on each side propelling it. For this work we focus on predicting the thrust/forward force generated by the tandem fins, both the cyclic average thrust and the profile of the time-history of the force generation. The shape of the time history is important towards understanding the nature of the flapping kinematics and it also holds clues towards fluid dynamic events such as fluid vortex separation and recapture around the fins that can significantly enhance or retard overall thrust generation. A major challenge in the training of the neural networks is the ability to learn these fluid dynamic events without any explicit programming.

\begin{figure}[htbp]
\centering \hspace{0in}
\mbox{\subfigure[Rectangular fin\label{fig-geo-rect}]{\includegraphics[height=1.5in,keepaspectratio]{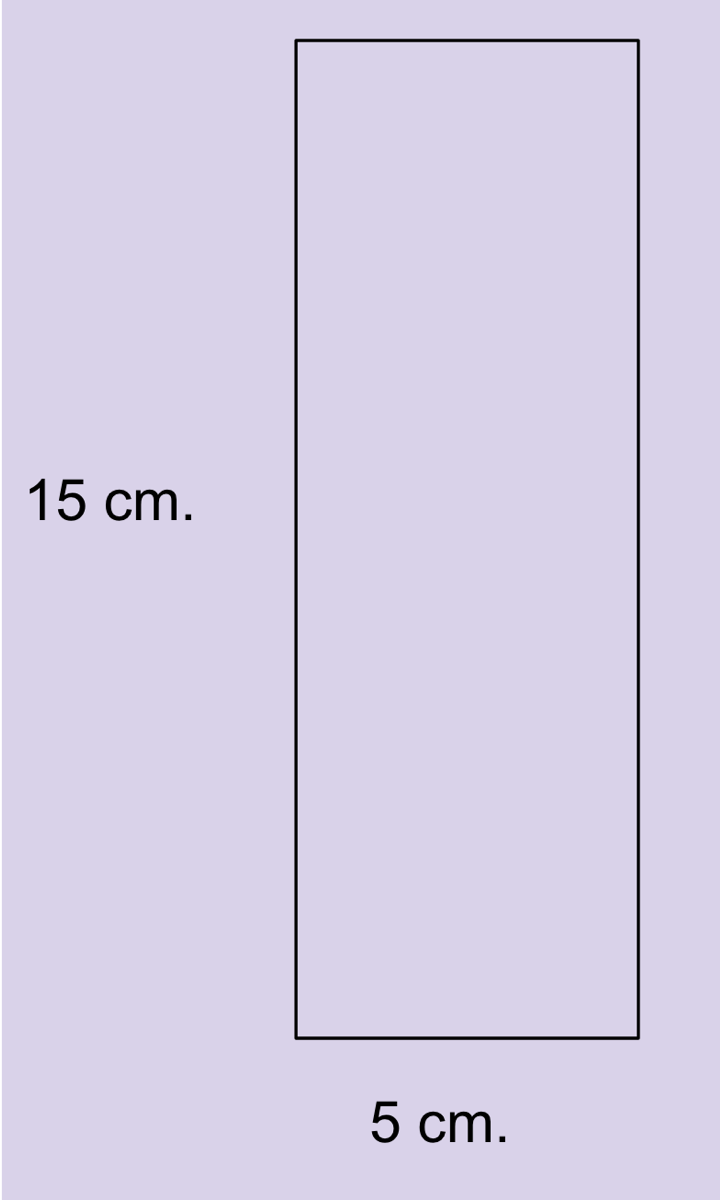}}}
\hspace{0.in}
\mbox{\subfigure[Bio-inspired fin\label{fig-geo-bio0}]{\includegraphics[height=1.5in,keepaspectratio]{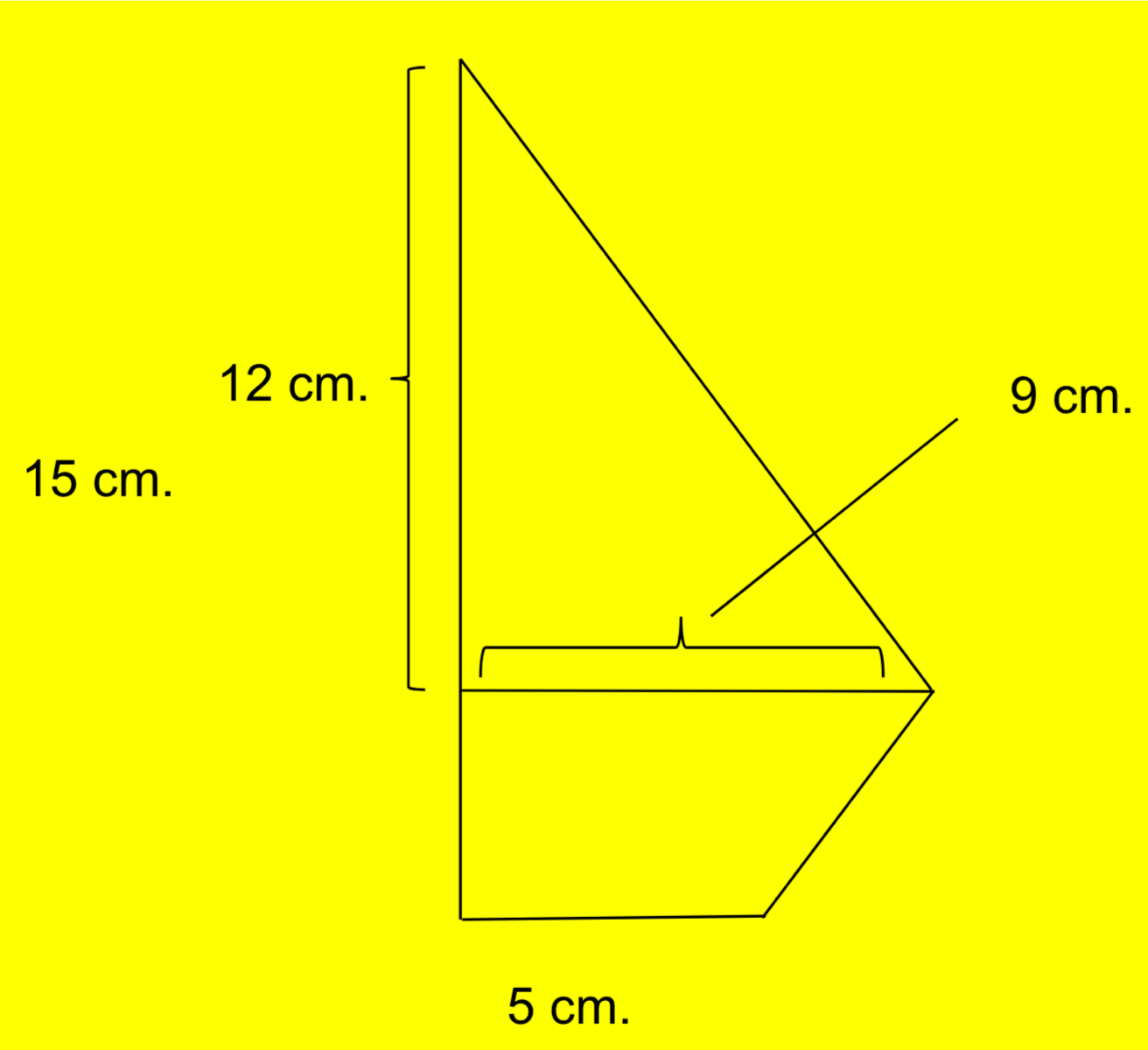}}}
\caption{Fin geometries used in experimental and CFD data collection.}
\label{fig-geometry}
\end{figure}

\begin{figure}[htbp]
\centering \hspace{0in}
\mbox{\subfigure[Bio-fin 1\label{fig-geo-bio1}]{\includegraphics[height=1.5in,width=1.5in]{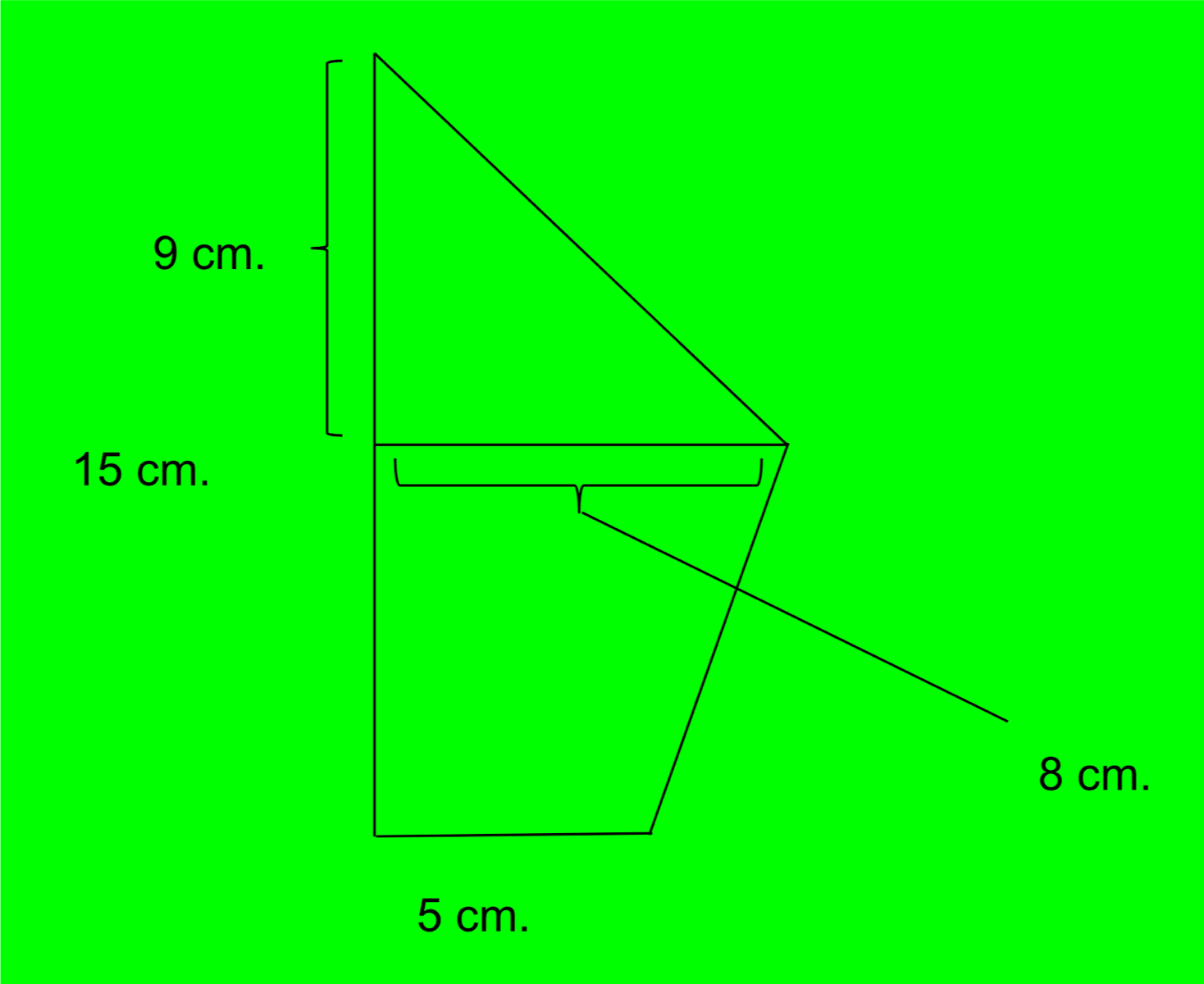}}}
\hspace{0.in}
\mbox{\subfigure[Bio-fin 2\label{fig-geo-bio2}]{\includegraphics[height=1.5in,width=1.5in]{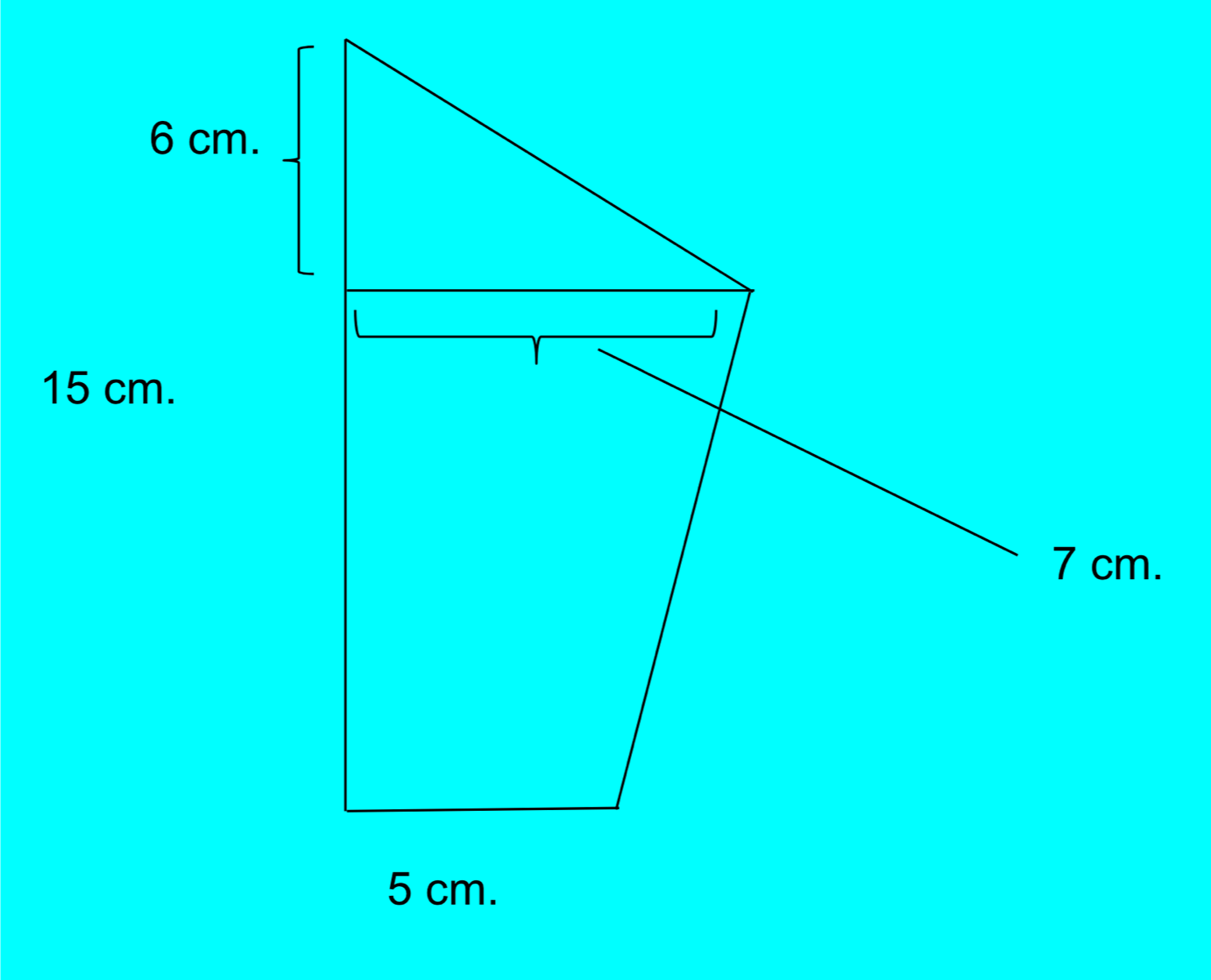}}}
\mbox{\subfigure[Bio-fin 3\label{fig-geo-bio3}]{\includegraphics[height=1.5in,width=1.5in]{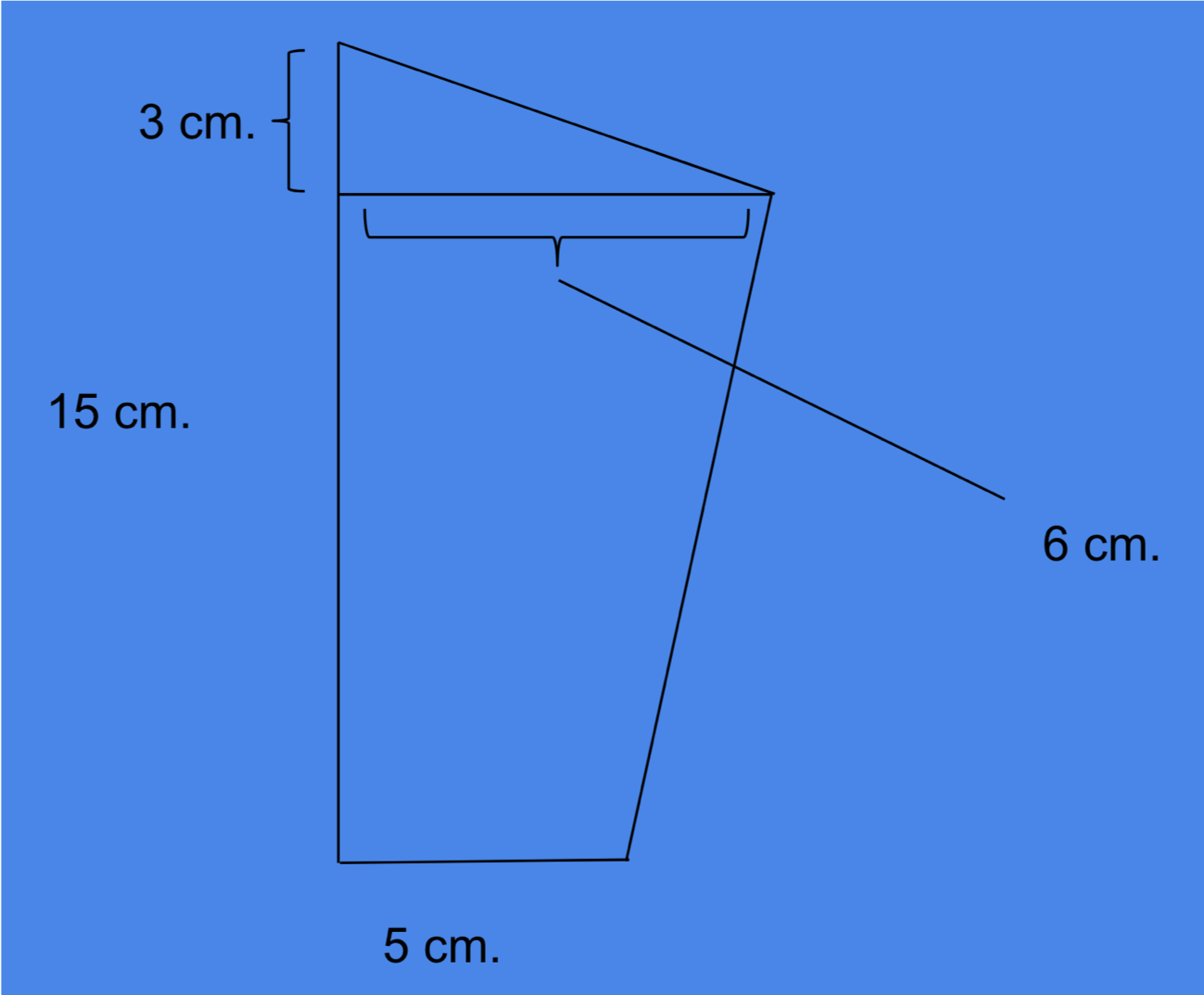}}}
\caption{Bio-inspired geometries that transition from bio-fin to rectangular fin as the test set.}
\label{fig-geometry2}
\end{figure}

In this work, all training set data are from two fin shapes, a rectangular fin (fig.~\ref{fig-geo-rect}) and a bio-inspired fin or bio-fin (fig.~\ref{fig-geo-bio0}). Tandem fin configuration experimental and CFD results for these fins have been published previously\cite{geder_underwater_2017,Ged18} for various flow regimes, fin configurations, and driven kinematics. We test the accuracy of the surrogate models for new fin shapes, fig.~\ref{fig-geometry2}, that  lie in-between the bio-fin and the rectangular fin, based on their ability to predict fluid dynamic effects in the thrust profile for these previously unseen geometries and compute the average thrust force.

Surrogate models are used extensively in engineering design due to the prohibitive cost of high-fidelity simulations that are an impractical tool to effectively search the design space. There are a number of surrogate modeling approaches, including reduced order models and lookup tables. However, machine learning approaches---approaches in which the model learns from data as opposed to being explicitly programmed---have become increasingly popular due to the increasing amount of available data and the rapid theoretical advances in the field. Machine learning approaches to surrogate modeling include nonlinear regression, tree ensembles, and kernel-based interpolations methods such as kriging.

More recently, deep learning approaches have been introduced as powerful modeling tool. Deep neural networks are hierarchical networks of computation units, or "neurons", that are capable of approximating arbitrary functions \cite{Hor89}. Deep learning has had a transformative effect in fields such as computer vision \cite{Kri12} and natural language processing \cite{Mik13a}, and it has shown promising results in physical modeling and design \cite{Rai19,Tri18a,Zhu18a}. Deep learning algorithms have several major strengths: they can represent arbitrarily complex functions; they can learn incrementally, reducing the memory cost of training and allowing for on-going model refinement; and their flexible structure and training process allows for easier sensitivity testing and analysis.

In this paper, we investigate the effectiveness of several deep learning surrogate models for fin geometry and configuration in multi-fin flapping propulsion systems.

The major contributions of this research are
\begin{enumerate}
  \item a new surrogate model for predicting the force profiles of novel fin geometries in multi-fin flapping propulsion systems, and
  \item a demonstration of the potential of neural-network-based surrogate modeling for propulsion system design.
\end{enumerate}

We hope that this work is a step towards a robust, fast surrogate model that is able to explore many dimensions of the design space.


\section{Methods}

In addition to a nonlinear regression baseline, we test five increasingly complex neural networks to predict the thrust given the propulsion system kinematics and geometry. Initial models are trained only on the lead fin data and their predictive accuracy is used to refine the input parameters and hyperparameters for the various models. The performance of the lead fin models is used to inform the setup and training of the rear fin neural network model. The architectures, training setups, and evaluation criteria of each model are described in the following sections.

\subsection{Training Dataset}

\subsubsection{Input Parameters}

The first problem to tackle is choosing the inputs that can effectively span the high dimensional space of the problem, the thrust being a function of many parameters. For the configuration shown in fig.~\ref{fig-tandemfin}, we have an input space that consists of several variables as described in table~\ref{tab-inputs}.

\begin{table}[htbp]
\caption{Training Parameters}
\begin{center}
\begin{tabular}{|c|c|}
\hline
\textbf{Parameters}   &   \textbf{Description} \\
\hline
\textbf{\textit{Geometry}}    & \\
\cline{1-1}
5 or more geometric                 &  At equidistant points  \\
chord lengths $^{\mathrm{a}}$ &  in the spanwise direction \\
\hline
Fin leading edge length (meters) & Reference length for normalization \\
\hline
$X_\text{offset}$ & Lead fin to rear fin offset distance \\
\hline
\textbf{\textit{Kinematics}}    & \\
\cline{1-1}
T (secs) & Time period of a fin stroke(flap) cycle \\
\hline
Stroke angle(radians) & Time varying history of the flapping \\
             & angle over a single cycle \\
\hline
Pitch angle(radians) & Time varying history of the pitching \\
            &  angle over a single cycle \\
\hline
Forward speed(m/s) & Incoming fluid flow velocity \\
\hline
Fin tip average speed(m/s) & Average speed of the \\
               & wing tip over a single cycle \\
\hline
Stroke phase offset & Phase offset of the rear  \\
         & fin flapping w.r.t to lead fin \\
\hline
Pitch phase offset & Phase offset of rear fin \\
        & pitching w.r.t to lead fin \\
\hline
\textbf{\textit{Forces}}    & \\
\cline{1-1}
Force (N) & Time varying profile of the \\
    & thrust over a single cycle \\
\hline
\multicolumn{2}{l}{$^{\mathrm{a}}$As shown in fig.~\ref{fig-geo-chord}.}
\end{tabular}
\label{tab-inputs}
\end{center}
\end{table}

The data from CFD and experimental runs are preprocessed to normalize input length scales, use force coefficient representation, match timescales, and remove outliers. The forward speed is normalized with the average fin tip speed. The chord lengths data, normalized with the leading-edge length, are constant for each fin geometry. Stroke and pitch angles are intended to give the spatial orientation of the fin over a cycle and are time dependent.

\begin{figure}[htbp]
\centerline{\includegraphics[height=1.5in,keepaspectratio]{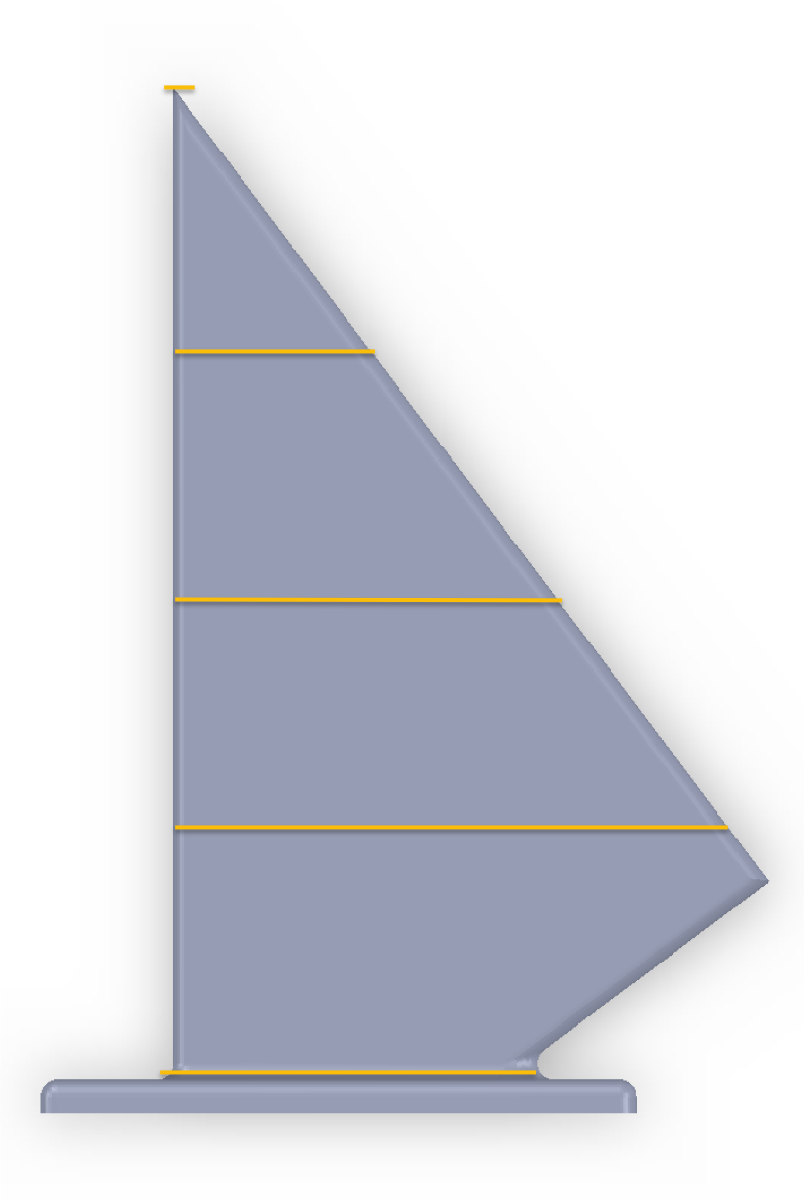}}
\caption{Fin geometry with chord lengths at 5 equidistant points.}
\label{fig-geo-chord}
\end{figure}

\subsection{Neural Network Architectures}

We considered two major architectures during these experiments:
densely-connected neural networks (DNNs) and 1D convolutional neural networks
(CNNs).

\begin{figure}[htbp]
\centerline{\includegraphics[height=1.5in,keepaspectratio]{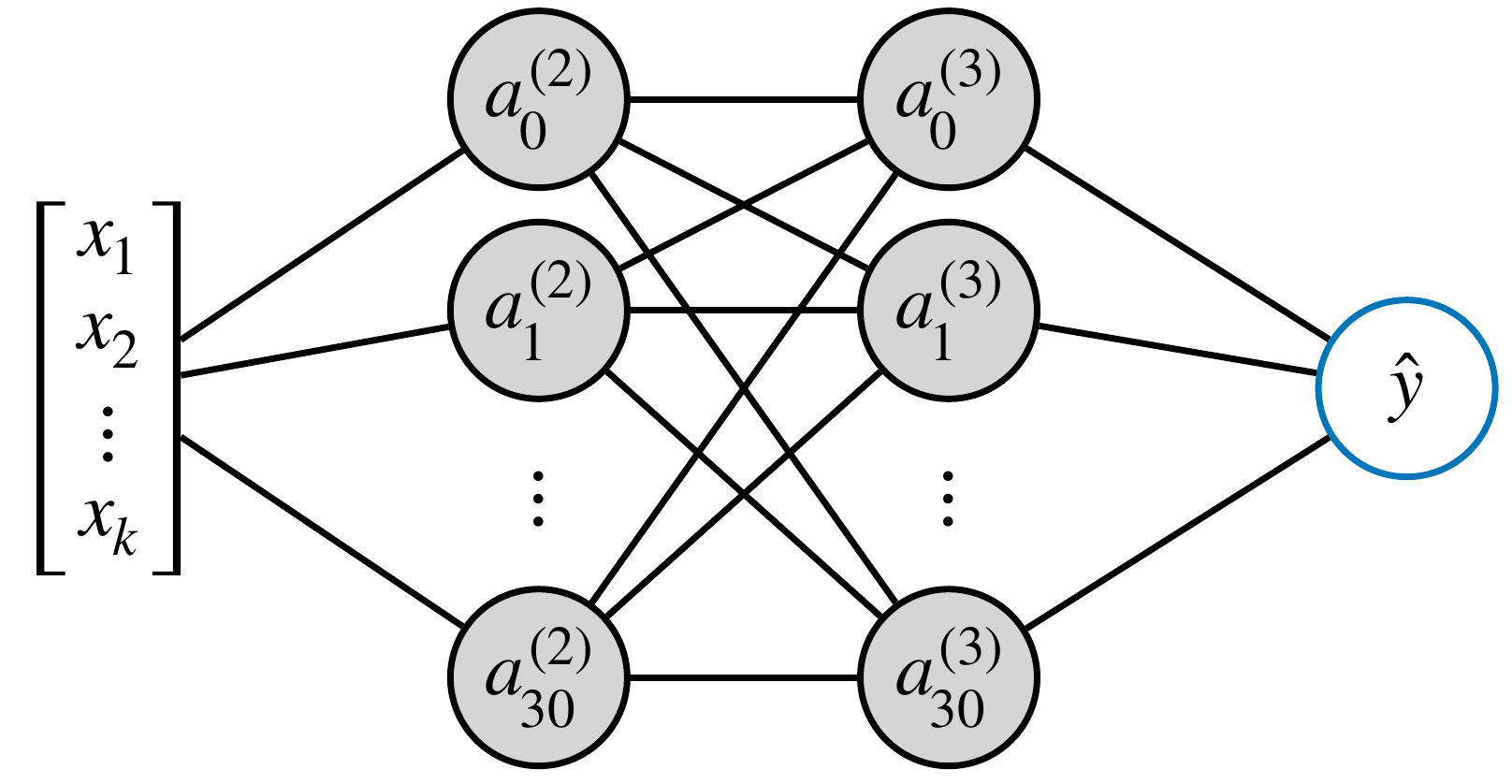}}
\caption{A basic densely-connected network. Each neuron is connected to all neurons in previous and subsequent layers.}
\label{fig-dnn}
\end{figure}

A DNN is a feed-forward neural network in which each neuron is connected to all neurons in the previous and subsequent layers as seen in fig.\ref{fig-dnn}. Training a DNN means learning a set of weight matrices $\Theta^{(l)}$ mapping each neuron in layer $l-1$ to each neuron in layer $l$. We use this as a baseline neural model because the dense connections mean that DNNs treat all data equally: they make no assumptions about the underlying structure of the data. Case B uses this basic architecture.

\begin{figure}[htbp]
\centerline{\includegraphics[width=\linewidth,keepaspectratio]{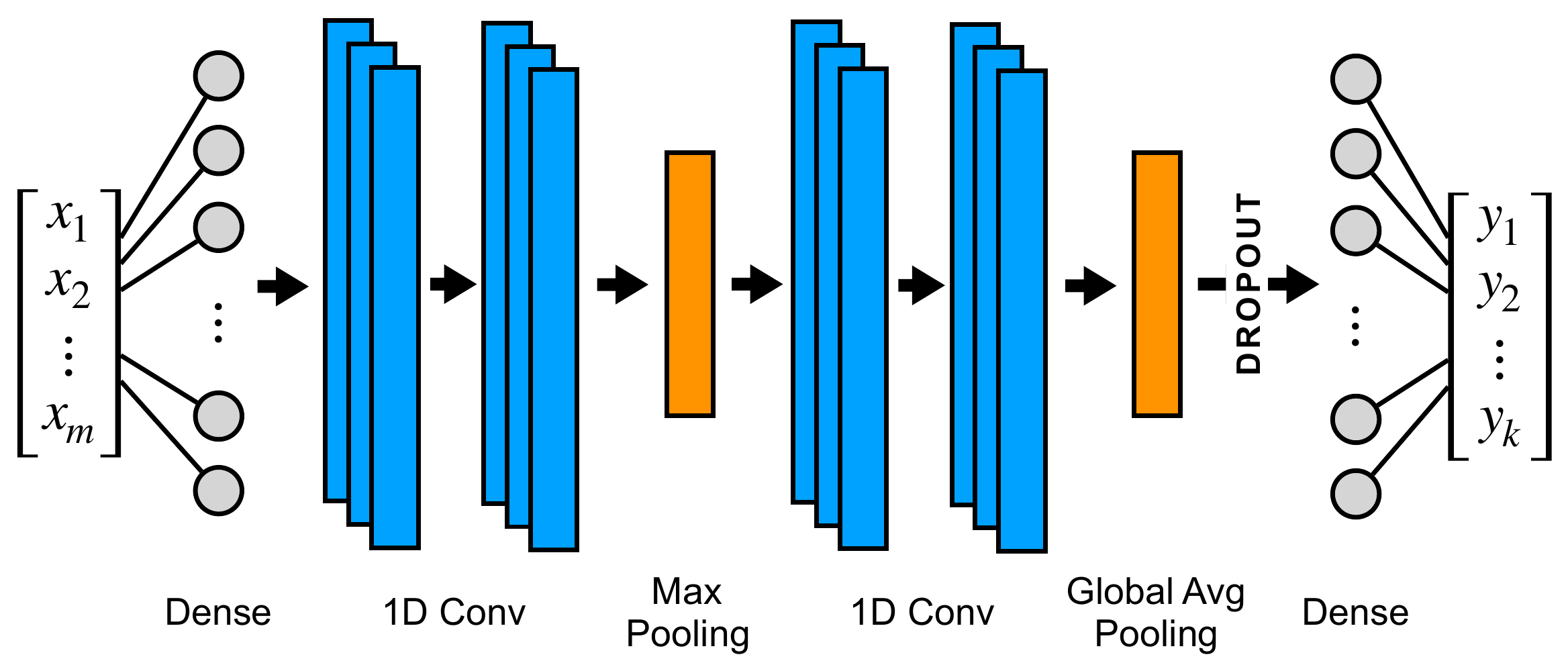}}
  \caption{A 1D convolutional neural network (CNN) with pooling layers and dropout regularization. Instead of learning parameters relating all neurons in adjoining layers, a convolutional neural network learns a smaller kernel of shared parameters.}
\label{fig-cnn}
\end{figure}

CNNs, in contrast, make assumptions about the data. Instead of learning the full weight matrices, CNNs learn a smaller set of shared parameters called a kernel for each layer (fig.~\ref{fig-cnn}). This parameter sharing enforces translational invariance and restricts the solution space by reducing the number of trainable parameters. As a result, it tends to be better than DNNs at capturing local structure.

In our convolutional models (cases C-E), we use the basic architecture shown in fig.~\ref{fig-cnn}: a seven layer, feed-forward architecture consisting of an input layer, two 1D convolution layers, a max pooling layer, two more 1D convolution layers, a global average pooling layer, and finally a fully-connected output layer. Each convolutional layer uses a leaky ReLU activation function \cite{Maa13} with the inactive gradient set at $\alpha=0.05$. Desirable traits for the trained network include robustness over the range of variations of the input quantities and minimizing overfitting to the training data spectrum. To this end, the network utilizes dropout regularization during training that randomly turns off some of the connections while training and steers the network away from learning an overly complex function, with 20\% of nodes being dropped per iteration \cite{Sri14}. All kernels are 3 channels deep, and the width varies per experiment.

\subsection{Training Setup}

The goal of training is to learn the mapping between parameters of the
flapping-fin configuration(see table~\ref{tab-inputs}) and the resulting force profile by minimizing a loss
function $\mathcal{L}$.

\begin{equation}
  \min_{\theta} \mathcal{L}(y,\hat{y})
  \label{eq-training}
\end{equation}
where $\theta$ are the trainable parameters of the network, $\hat{y}$ is the
predicted force profile and $y$ is the true force profile for a given input.
All runs used the mean squared error as the loss function unless noted (eq.\ref{eq-mse}).
\begin{equation}
  \mathcal{L}_\text{MSE} = \frac{1}{N} \sum_{i=1}^{N} \left(y_i - \hat{y}_i\right)^2
  \label{eq-mse}
\end{equation}

Networks were tuned using a grid search (as seen in table \ref{tab-hyperparam0}). The network architecture and optimization hyperparameters were varied along the following axes: number of layers, number of units/layer, activation function, loss function, learning rate, and data granularity. Due to the high cost of collecting and preparing training data, this last value, data efficiency, was of particular interest.
All networks were trained for 1,500 epochs with a batch size of 500 examples. Networks were optimized used the Adam \cite{Kin15a} optimizer using decay rates $\beta_1=0.9$ and $\beta_2=0.999$ and a learning rate of $\alpha=0.01$.
All networks were trained using Keras with a TensorFlow \cite{tensorflow} backend.

\begin{table}[htbp]
\caption{Hyperparameters}
\begin{center}
\begin{tabular}{|c|c|c|}
\hline
\textbf{Number of Layers}& \multicolumn{2}{c|}{\textbf{2}}   \\
\hline
\textbf{Number of units}& 16 & 32 \\
\hline
 \textbf{Activation Function}& \textbf{TanH} & \textbf{Leaky ReLU} \\
 \hline
 \textbf{Loss Function}& \multicolumn{2}{c|}{\textbf{MSE}} \\
\hline
 \textbf{Learning Rate}& 0.001 & 0.003 \\
 \hline
 \textbf{Data Granularity}& \multicolumn{2}{c|}{1000 or more discrete pts. per cycle}\\
\hline
\end{tabular}
\label{tab-hyperparam0}
\end{center}
\end{table}

The processed dataset was split randomly into training and test datasets of 85\% and 15\% of the data respectively. For reproducability and consistency between runs, the random seed was set to 8 for all runs.

\subsection{Evaluation Criteria}

The objective of our surrogate models is to quickly and accurately predict the thrust profile of a novel multi-fin configuration. To achieve this, we evaluate models on empirical error, generalizability, data efficiency, and computational performance.

Empirical error, or the correctness of our predictions, is our primary evaluation criteria. This is measured as mean squared error of our prediction versus the expected value as seen in eq.~\ref{eq-mse}. We also verify if the cycle averaged thrust predictions lie within the experimental error bounds of $\pm 0.047~\text{N}$ for the lead fin and $\pm 0.050~\text{N}$ for the rear fin.

To measure generalizability, or the model's ability to make accurate predictions for unfamiliar fin geometries, we validate the models' performance on two biologically-inspired geometries that were not part of the training set.

Data efficiency, or the network's ability to learn an accurate model given limited training data, is evaluated during the hyperparameter tuning process. Models are trained for different fractions of the training data, and the final selected models are able to produce accurate predictions given varying subsets of the full training set.

Finally, we measure computational performance by tracking the training time and benchmarking prediction time of five fin stroke cycles. This is then compared against the time requirements of experimental trials and traditional CFD modeling.

\section{Results and Discussion}

\subsection{Case A:Nonlinear Regression}

\begin{figure}[htbp]
\centerline{\includegraphics[height=2.in,keepaspectratio]{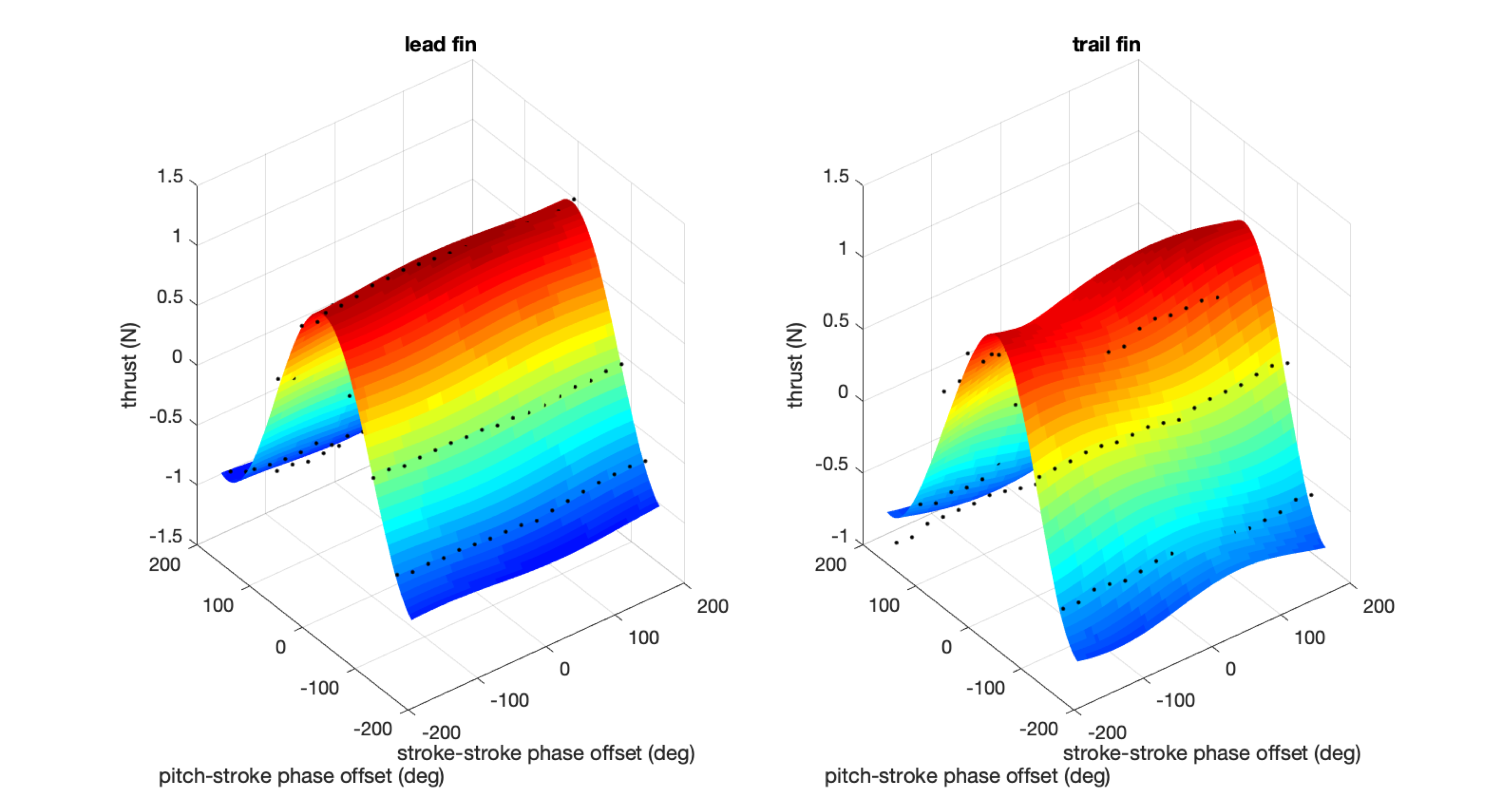}}
\caption{Tandem pectoral fin thrust results as a function of stroke phase offset between fins and pitch-stroke reversal offset within an individual fin fitted with cross-correlated harmonic functions..}
\label{fig-nlreg}
\end{figure}

An initial approach using nonlinear regression is implemented to model cycle-averaged thrust as a function of pitch-stroke phase offset within individual fins and stroke-stroke phase offset between fins.  Using a cross-correlated harmonic function (eq.~\ref{eq-nlreg}), predictions of stroke-averaged thrust for the lead and rear fins have MSE = $0.0083~\text{N}^2$ and MSE = $0.0074~\text{N}^2$, respectively, where average thrust for each fin ranges between -1.1 N and 1.1 N (fig.~\ref{fig-nlreg}).  While this method produced a reasonably good estimate for stroke-averaged thrust across a limited number of variables, understanding the effects of more variables on the time-varying thrust profiles will be prohibitively complex for this methodology.
\begin{equation}
  F_{T} = a_1 + a_2 sin(x + a_3y + a_4) + a_5sin(y + a_6x + a_7)
  \label{eq-nlreg}
\end{equation}
where $x$ is the stroke-stroke phase offset (between fins), and $y$ is the pitch-stroke phase offset (single fin), and $a_1-a_7$ are coefficients being solved for.

\subsection{Case B: DNN Baseline}

The goal of this first experiment is to establish a baseline for a neural network surrogate model mapping kinematics to the force profile. As a precursor, we explored what a DNN network could learn with minimal input parameters, like time, shape, spacing, and offset of the fins by training a large number of models with varying hyperparameters like number of hidden layers, units, learning rate, and activation functions. The best training convergence achieved is with a 3 layer densely connected network with the relevant hyperparameters listed in table~\ref{tab-caseB-hyp} and the inputs listed in table~\ref{tab-caseB}. The learning rate during the training varied between 0.001 to 0.003 and the training data included the lead fin data of both fin shapes, bio-inspired and rectangular.

\begin{table}[htbp]
\caption{Case B: Parameters}
\begin{center}
\begin{tabular}{|c|c|}
\hline
\multicolumn{1}{|c|}{\textbf{Input Parameters}}& \textbf{Output} \\
\cline{1-2}
\textbf{\textit{Lead fin}}&  \textbf{\textit{Predictions}} \\
\hline
Categorial shape of 0(bio-fin) and 1(rectangular fin)&   \\
\cline{1-1}
Stroke and pitch angle time history &  Force output  \\
\cline{1-1}
Forward speed&   \\
\hline
\end{tabular}
\label{tab-caseB}
\end{center}
\end{table}

\begin{table}[htbp]
\caption{Case B: Hyperparameters}
\begin{center}
\begin{tabular}{|c|c|}
\hline
\textbf{Number of Layers}& 3   \\
\hline
\textbf{Number of units}& 32 \\
\hline
 \textbf{Activation Function}& \textit{TanH}  \\
 \hline
 \textbf{Loss Function}& \textit{MSE} \\
\hline
\end{tabular}
\label{tab-caseB-hyp}
\end{center}
\end{table}

\begin{figure}[htbp]
\centering \hspace{0in}
\mbox{\subfigure[Rectangular fin\label{fig-mlp-rect}]{\includegraphics[height=2.in,width=3.in]{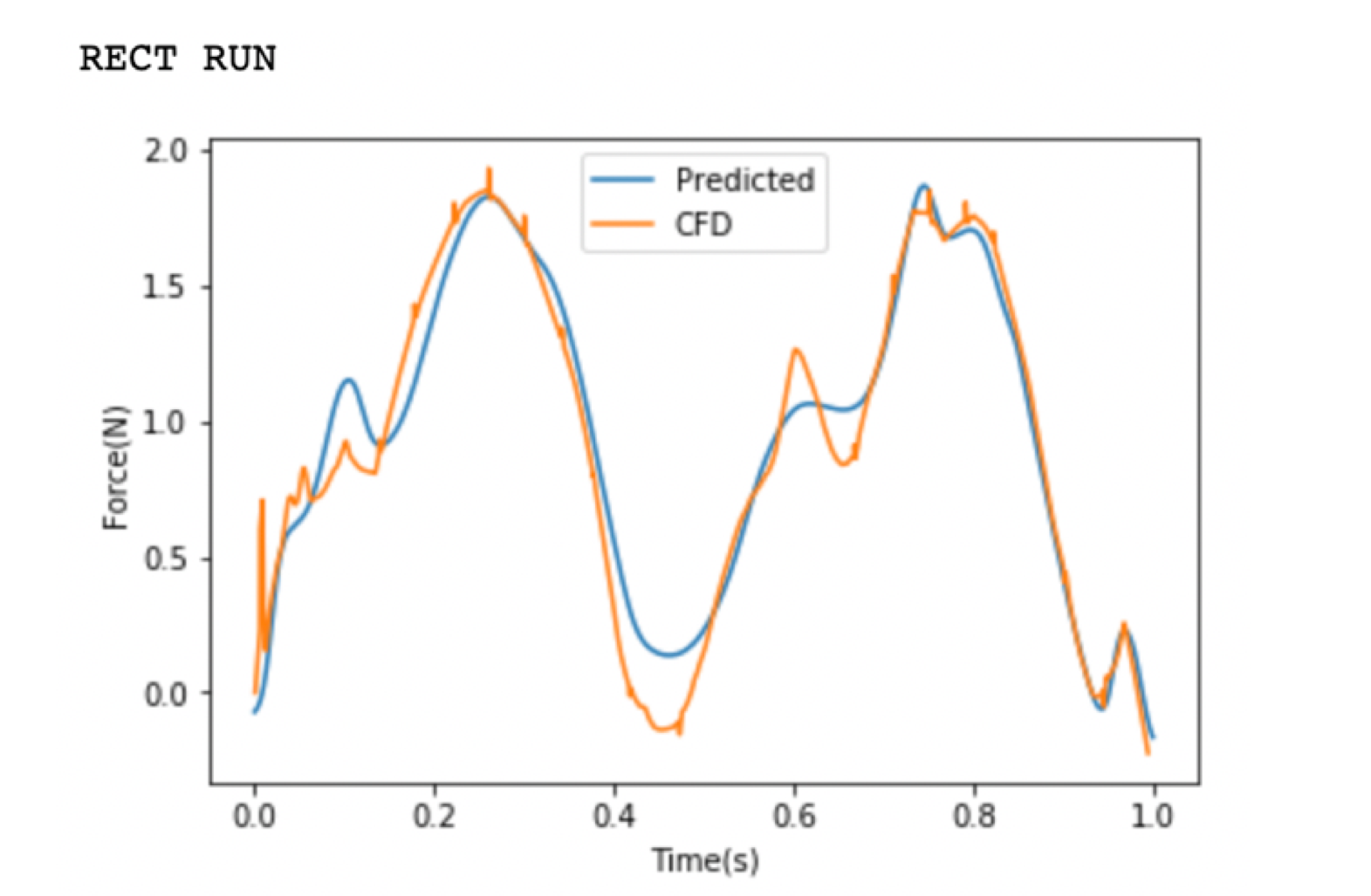}}}
\hspace{0.in}
\mbox{\subfigure[Bio-inspired fin\label{fig-mlp-bio0}]{\includegraphics[height=2.in,width=3.in]{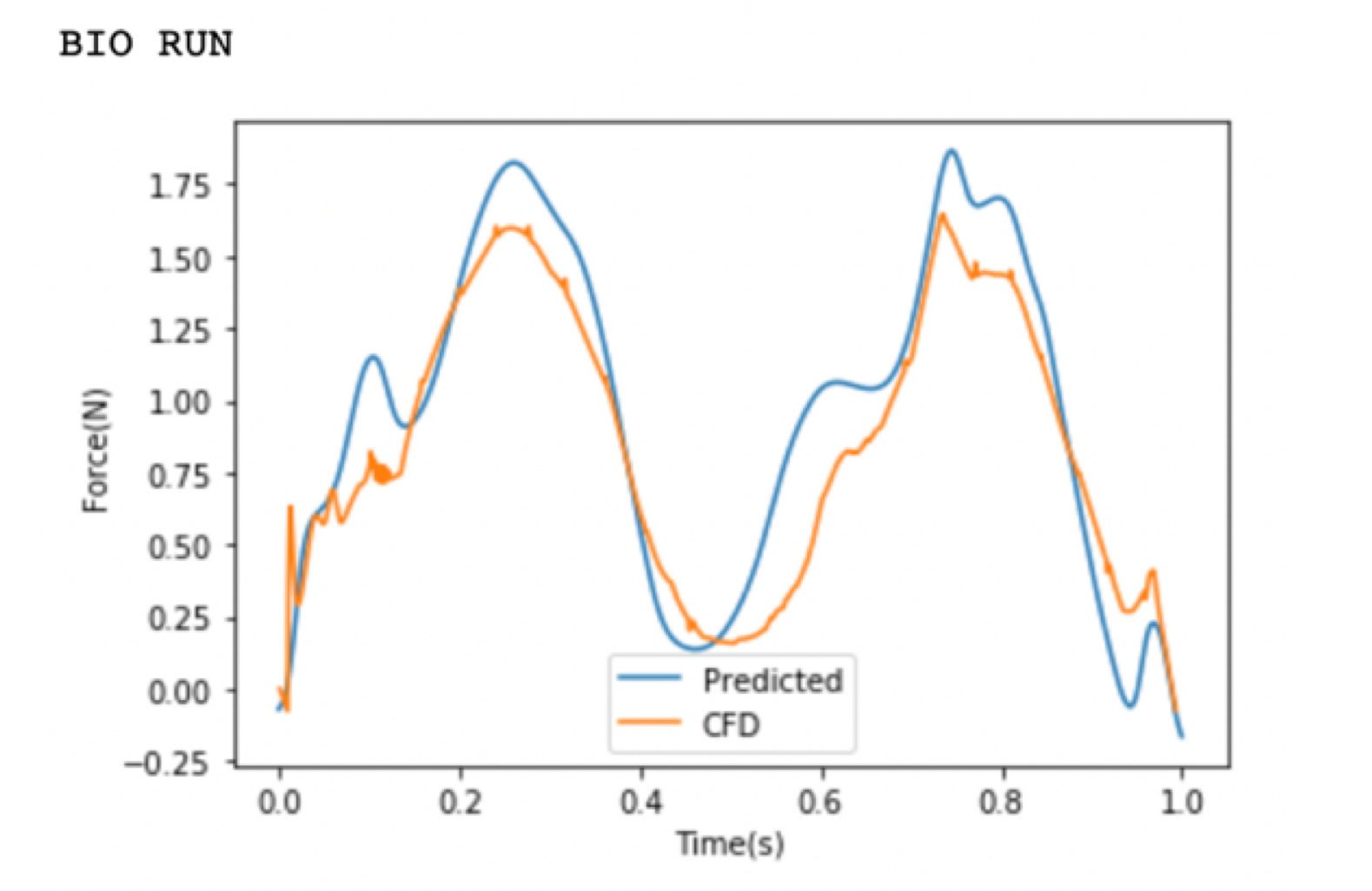}}}
\caption{DNN predictions.}
\label{fig-mlp}
\end{figure}

The model predictions fit for both fins shapes are shown in fig.~\ref{fig-mlp}. The model is able to learn some general relationship between the fin and its force output. However, with a MSE = $0.066~\text{N}^2$ and the force range being approximately 2 N, the predictions are clearly off in magnitude and do not capture local peaks too well, even though it exhibits subtle maxima.

\subsection{Case C: CNN with Spatial Inputs}

In order to better capture local structure, reduce overfitting, and improve data efficiency, the next experiment is a 1-D CNN architecture. As generalizability is a primary goal in this work, there is a need to develop an improved representation of the model geometry. In addition, we have to improve the capture of local peaks, that might be driven by fluid dynamic events such as vortex shedding or recapture, and are also dependent on fin geometry. To do this, we refine the input space, given in table~\ref{tab-caseC}, by providing five cross-sectional chord lengths, equi-spaced along the fin span from the root to the tip of the fin, to better represent the fin shape. The chord length data, normalized with the leading-edge length, are constant for each fin geometry. The time dependent stroke and pitch angle variation phase matched with output force during network training will impart an understanding of the spatial orientation of a given fin during the training. Filters from the 1-D CNN focus on a specific part of the sequence so that the network progressively reduces the size of the representation/number of parameters, making this more data efficient. The hyperparameters for this network are listed in table~\ref{tab-caseC-hyp}.

\begin{table}[htbp]
\caption{Case C: Parameters}
\begin{center}
\begin{tabular}{|c|c|}
\hline
\multicolumn{1}{|c|}{\textbf{Input Parameters}}& \textbf{Output} \\
\cline{1-2}
\textbf{\textit{Lead fin}}&  \textbf{\textit{Predictions}} \\
\hline
5 geometric chord lengths$^{\mathrm{a}}$&   \\
\cline{1-1}
Stroke and pitch angle time history &  Force output  \\
\cline{1-1}
Forward speed&   \\
\hline
\multicolumn{2}{l}{$^{\mathrm{a}}$As shown in fig.~\ref{fig-geo-chord}.}
\end{tabular}
\label{tab-caseC}
\end{center}
\end{table}

\begin{table}[htbp]
\caption{Case C: Hyperparameters}
\begin{center}
\begin{tabular}{|c|c|}
\hline
\textbf{Initial dense layer}& yes   \\
\hline
\textbf{Number of Layers}& 2   \\
\hline
\textbf{Number of units}& 16 \\
\hline
 \textbf{Activation Function}& \textit{TanH}  \\
 \hline
\end{tabular}
\label{tab-caseC-hyp}
\end{center}
\end{table}

\begin{figure}[htbp]
\centering \hspace{0in}
\mbox{\subfigure[Bio-inspired fin\label{fig-cnn0-bio0}]{\includegraphics[width=3.0in,keepaspectratio]{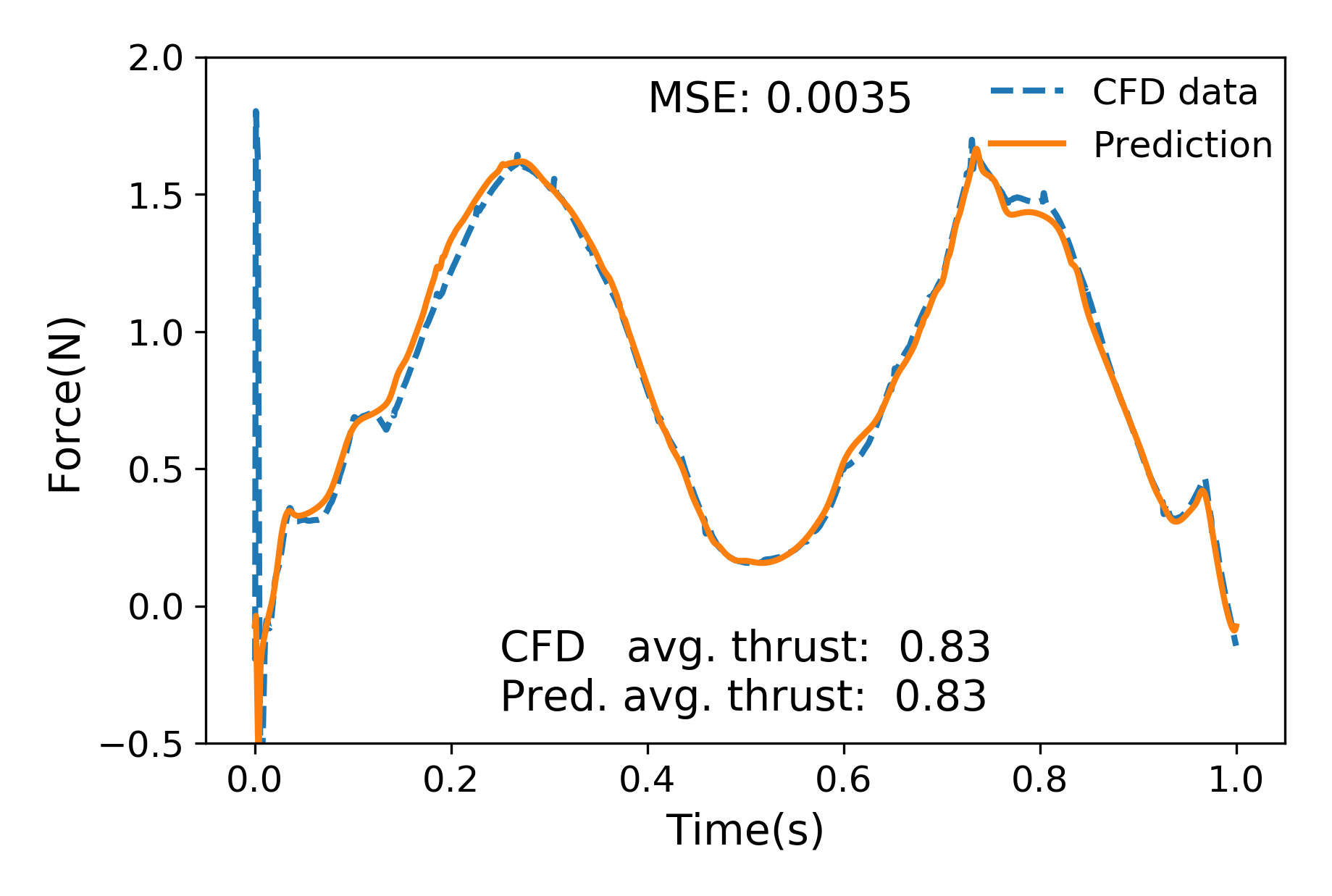}\hspace{0.35in}}}
\hspace{0.in}
\mbox{\subfigure[Rectangular fin\label{fig-cnn0-rect}]{\includegraphics[width=3.in,keepaspectratio]{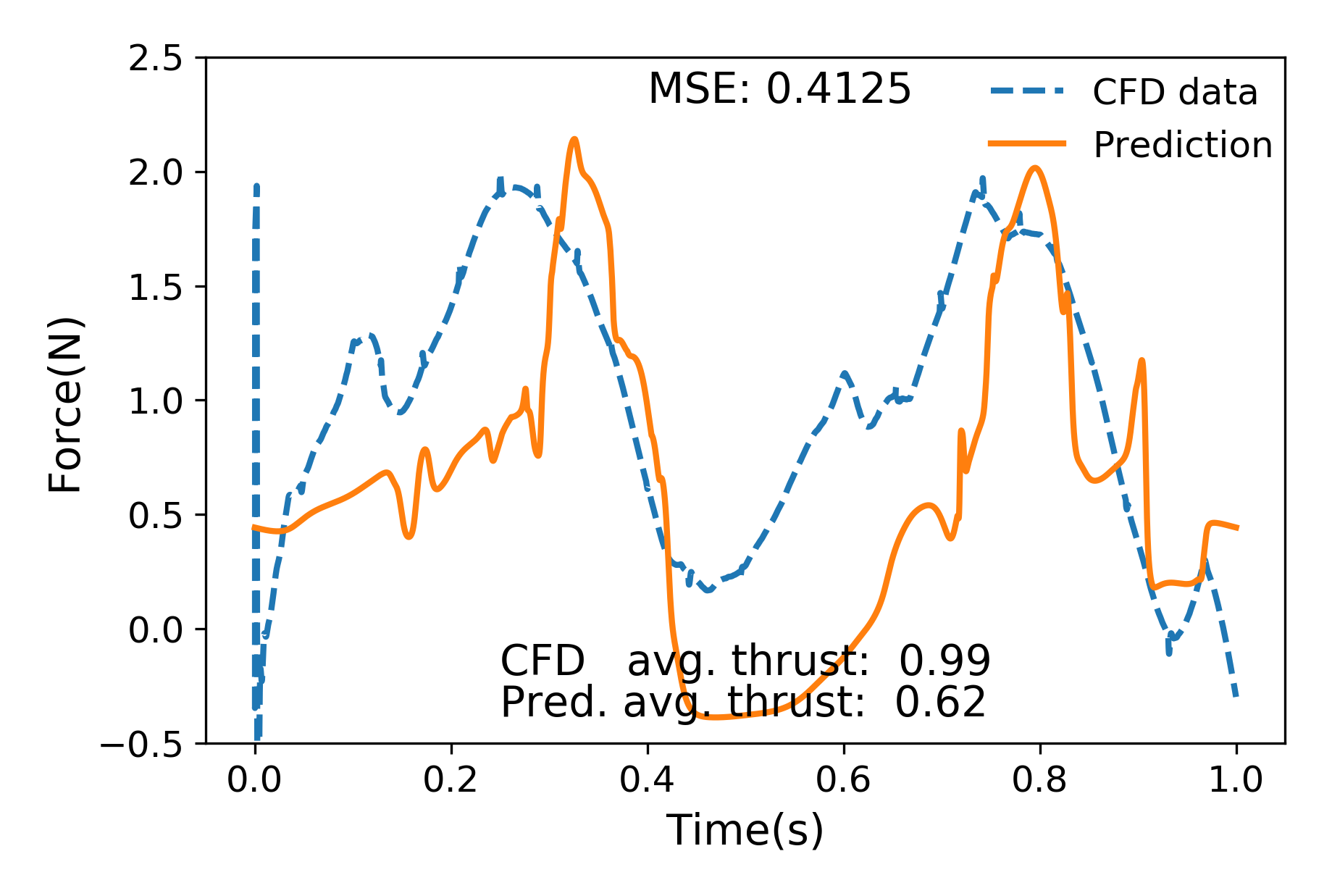}\hspace{0.35in}}}
\caption{Predictions from network trained only on bio-inspired fin data.}
\label{fig-cnn0}
\end{figure}

To test generalizability, in this experiment the network is trained only on the bio-inspired fin data and the trained network is used to predict the force output for the rectangular fin. Figure \ref{fig-cnn0} shows the network is able to get a good fit for the bio-inspired fin it trained on, with the cycle averaged thrust matching the CFD. The MSE = $0.0035~\text{N}^2$ implies, over the thrust cycle, the fit of the prediction with CFD data is within 4\% of the approximate peak to peak spread of 1.5 N. Interestingly, the network does seem able to predict events for the rectangular fin. The magnitudes of the force prediction on the rectangular fin show considerable discrepancies but it exhibits comparable peaks as the CFD data with a phase delay during the cycle, as seen in fig.~\ref{fig-cnn0-rect}. The CFD results clearly have a local vortex recapture sequence at stroke reversal in addition to the bigger peaks from the up and down strokes and the network weakly predicts these events. As expected, a single fin geometry is not sufficient training data for the network and therefore, for all subsequent cases we train on both fin shapes.

\subsection{Case D: Generalization to Unseen Geometries}

The main goal in the previous experiment, along with adding geometric data on the five chord lengths, was to enable the networks ability to generalize to a shape it has not seen.  For a surrogate model to be used as a design tool, the performance of the model on unseen geometries is a critical performance metric and we take another step towards this goal. This experiment tests the previous case's architecture on three new fin geometries that keep the area,75 sq. cm, constant and geometrically transitions from the bio-fin to the rectangular fin. Figure~\ref{fig-geometry2} shows the geometry and nomenclature of these new fins.

\begin{figure}[htbp]
\centering \hspace{0in}
\mbox{\subfigure[Bio-inspired fin]{\label{fig-cnn1-rect}{\includegraphics[width=2.5in,keepaspectratio]{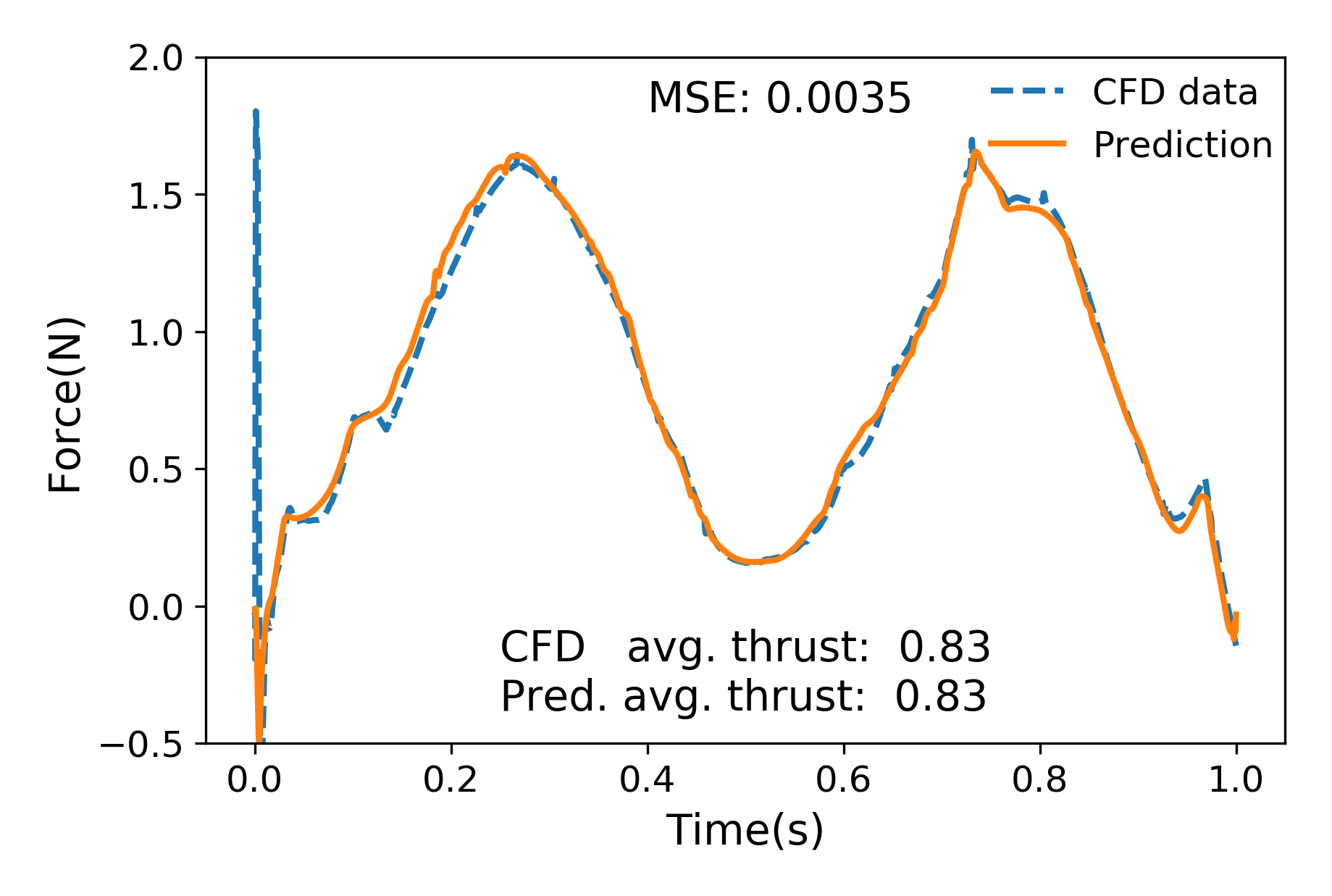}}
{\includegraphics[width=1.in,keepaspectratio]{geo-bio0.png}}}}
\hspace{0.in}
\mbox{\subfigure[Rectangular fin]{\label{fig-cnn1-bio0}{\includegraphics[width=2.5in,keepaspectratio]{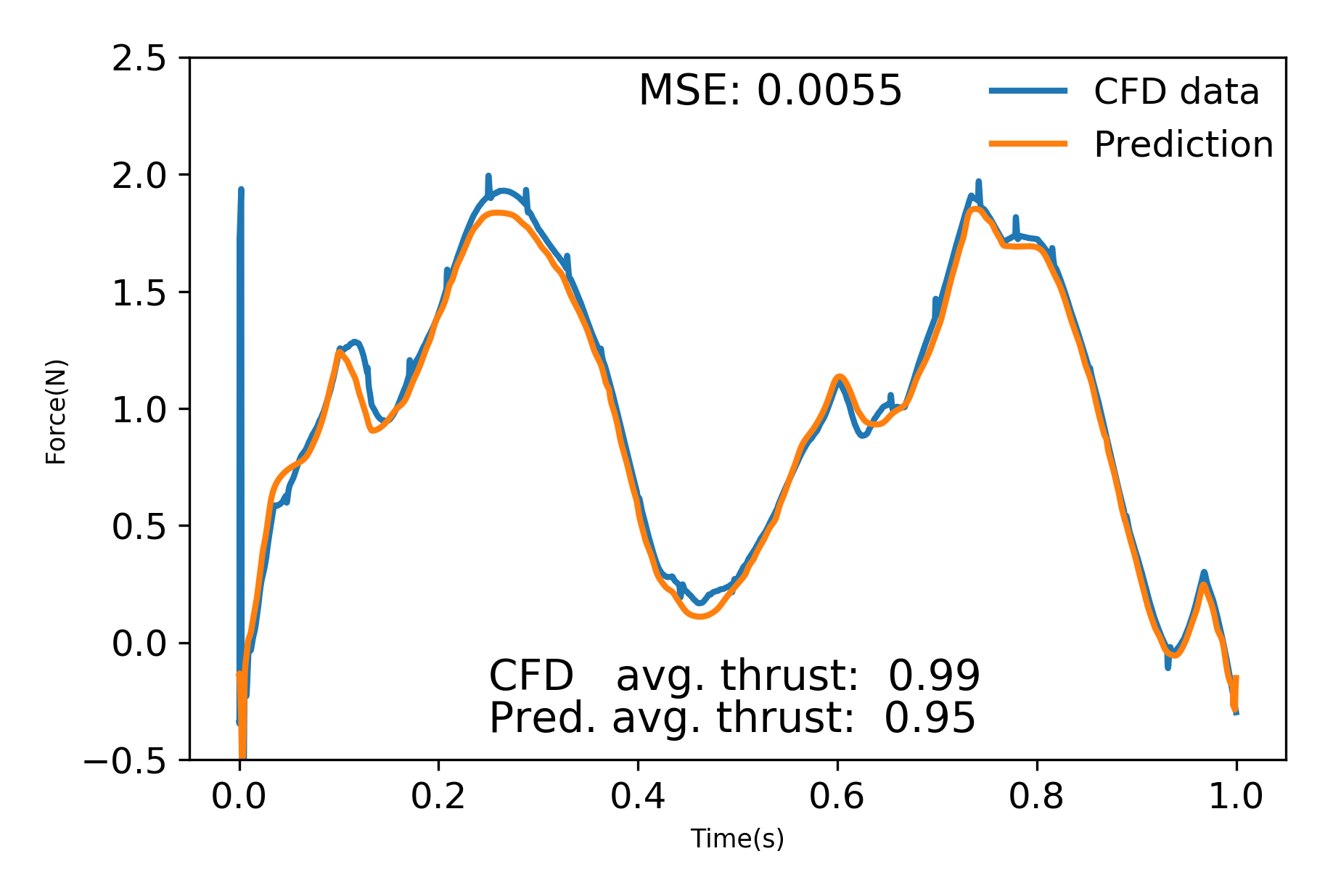}}
{\includegraphics[width=1.in,keepaspectratio]{geo-rect.png}}}}
\caption{5-point geometry network fit, after training on both fins.}
\label{fig-cnn1}
\end{figure}

\begin{figure}[htbp]
\centering \hspace{0in}
\mbox{\subfigure[Bio-fin 1 prediction]{\label{fig-cnn1-bio1}{\includegraphics[width=2.5in,keepaspectratio]{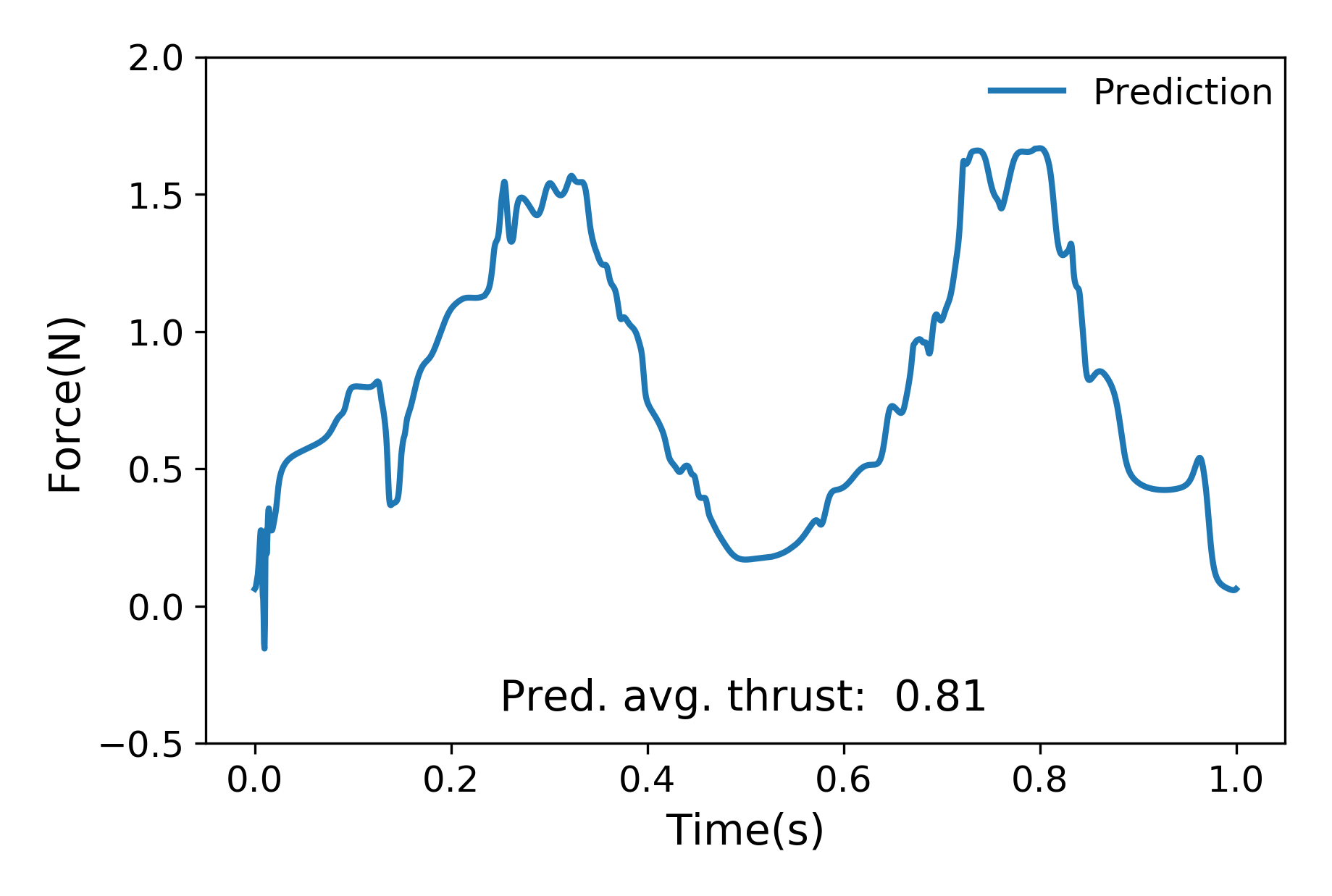}}
{\includegraphics[width=1.in,keepaspectratio]{geo-bio1.png}}}}
\hspace{0.in}
\mbox{\subfigure[Bio-fin 2 prediction]{\label{fig-cnn1-bio2}{\includegraphics[width=2.5in,keepaspectratio]{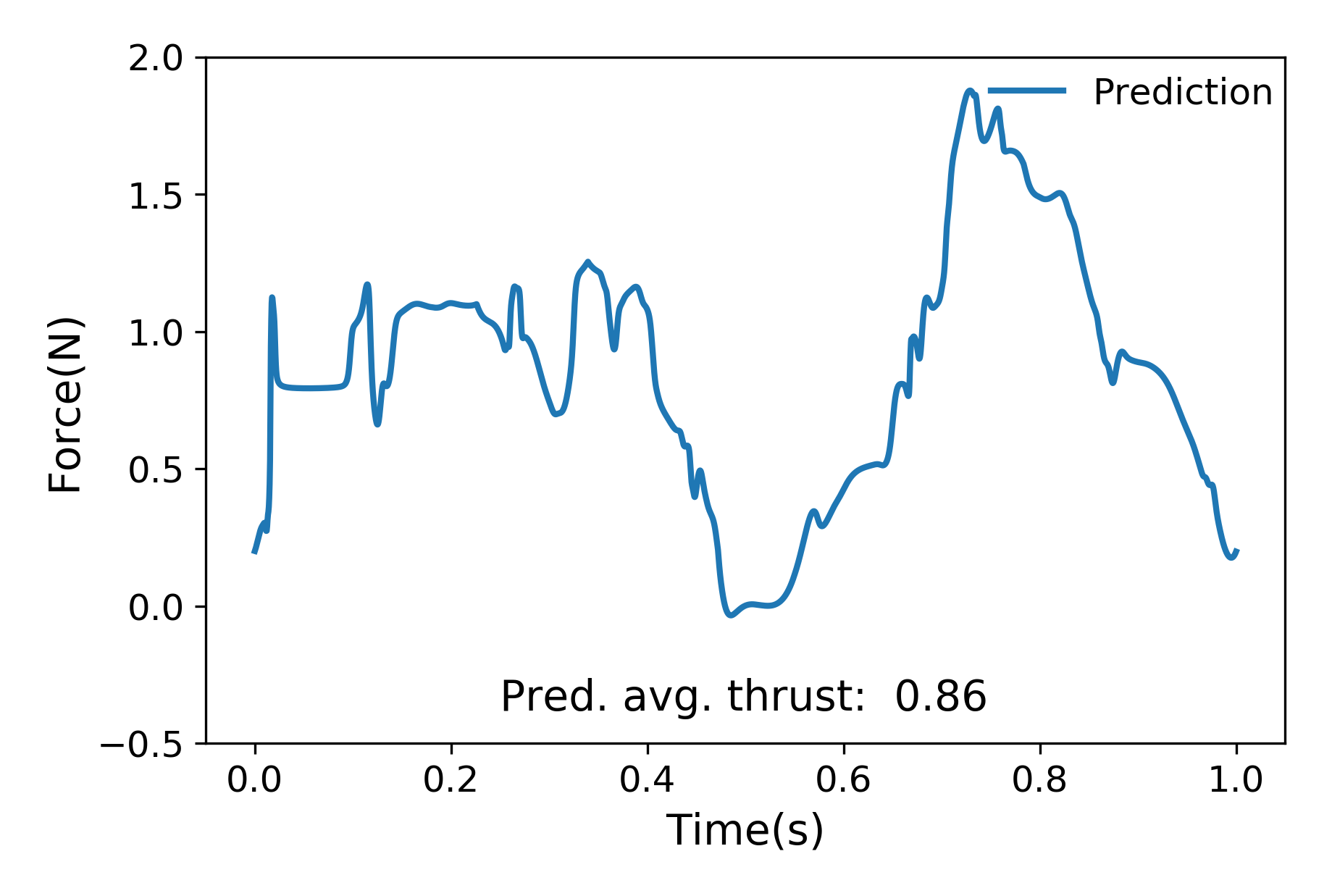}}
{\includegraphics[width=1.in,keepaspectratio]{geo-bio2.png}}}}
\mbox{\subfigure[Bio-fin 2 prediction]{\label{fig-cnn1-bio3}{\includegraphics[width=2.5in,keepaspectratio]{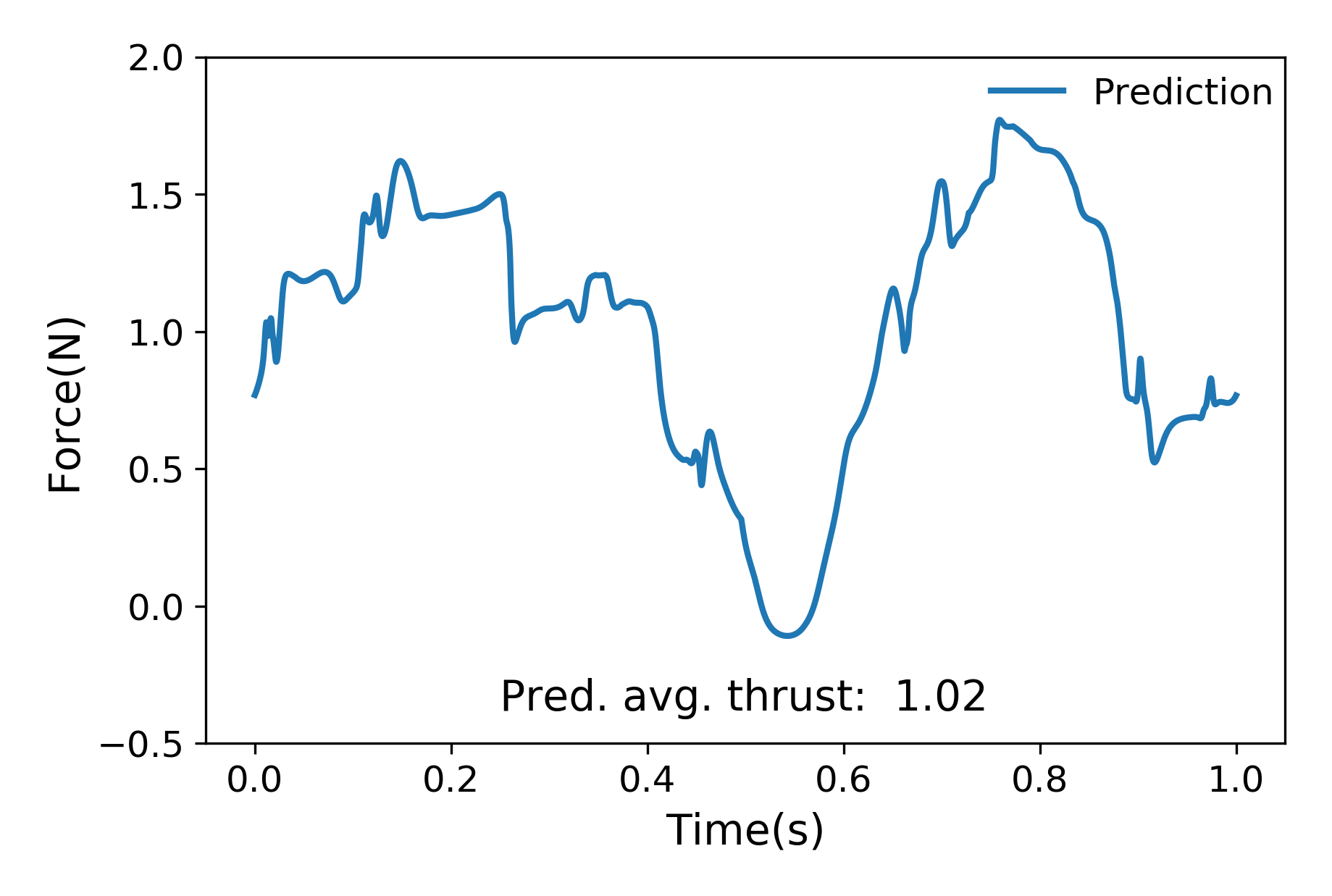}}
{\includegraphics[width=1.in,keepaspectratio]{geo-bio3.png}}}}
\caption{5-point geometry network predictions on different configurations.}
\label{fig-cnn1-pred1}
\end{figure}

Training is done using both fin shapes and fig.~\ref{fig-cnn1} shows that the network is able to learn the relation between the force profile and the shapes it trained on, with cycle averaged thrust of both fins lying within experimental bounds of $\pm 0.047~\text{N}$. Figure \ref{fig-cnn1-pred1} shows the network's force predictions on the new test geometries. Bio-fins 1(fig.~\ref{fig-cnn1-bio1}) and 3(fig.~\ref{fig-cnn1-bio3}) exhibit an expected thrust profile with peaks for stroke reversal and positive thrust during both upstroke and downstroke phases. Bio-fin 2 has the most departure with only one peak predicted, even though the predicted average thrust might seem reasonable. This outcome is not unexpected since bio-fin 2 is the furthest in geometry from the two shapes the network trained on.

\subsection{Case E: Refined Spatial Inputs}

\begin{figure}[htbp]
\centering \hspace{0in}
\mbox{\subfigure[5-point geometry comparison of bio-fin 2]{\label{fig-cnn1-bio2-2}{\includegraphics[width=2.8in,keepaspectratio]{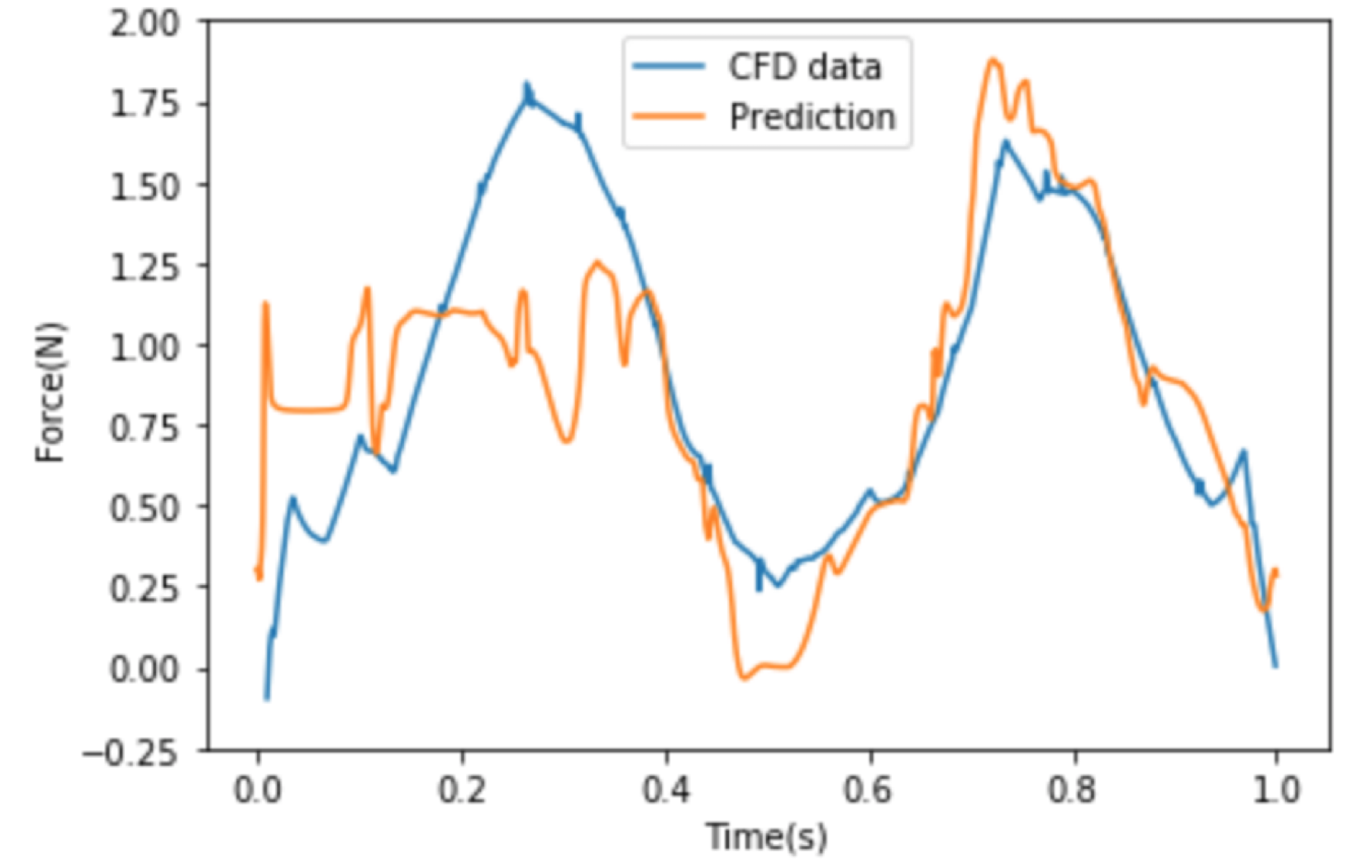}}}}
\hspace{0.in}
\mbox{\subfigure[10-point geometry comparison of bio-fin 2]{\label{fig-cnn2-bio2}{\includegraphics[width=2.8in,keepaspectratio]{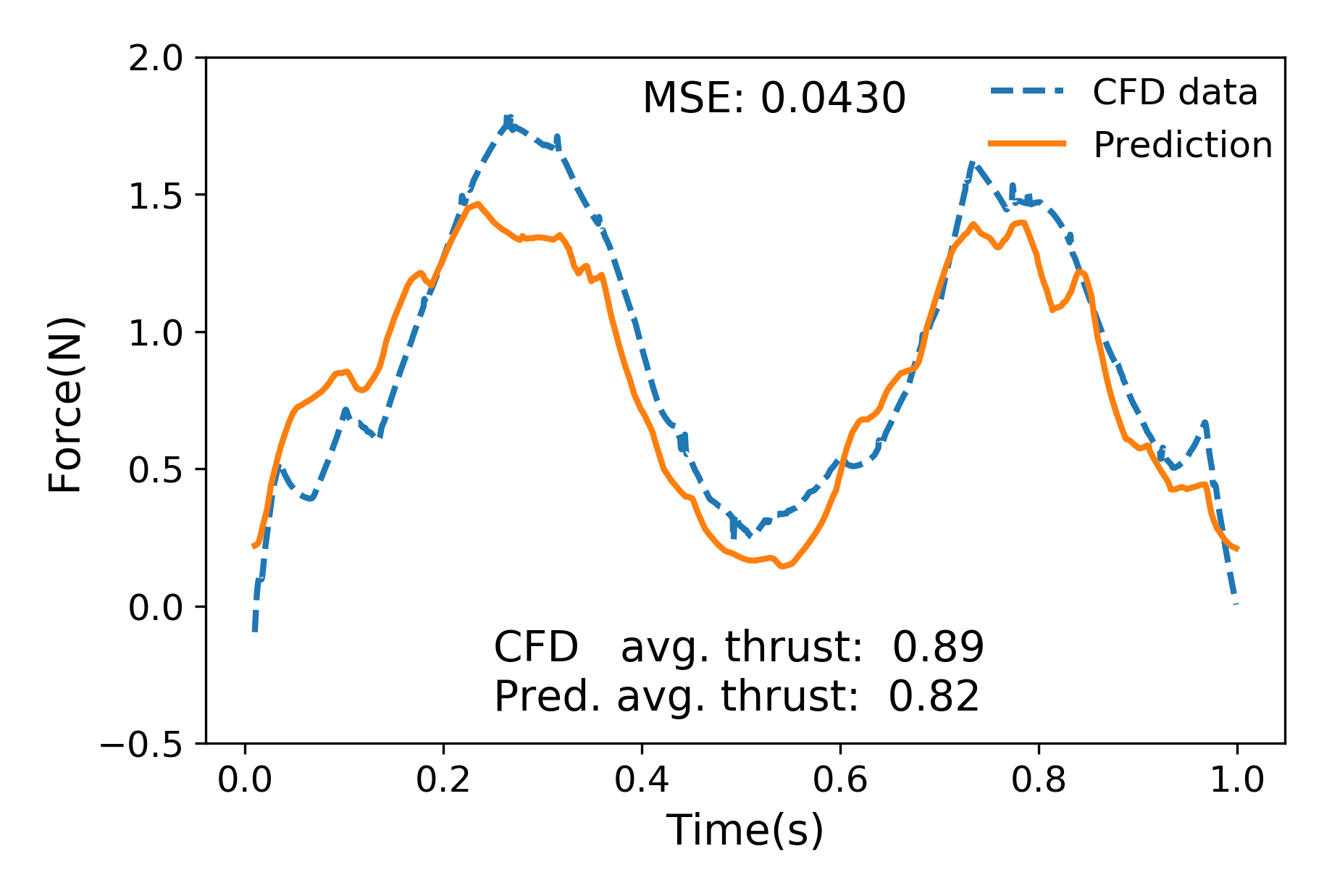}}}}
\mbox{\subfigure[10-point geometry comparison of bio-fin 1]{\label{fig-cnn2-bio1}{\includegraphics[width=2.8in,keepaspectratio]{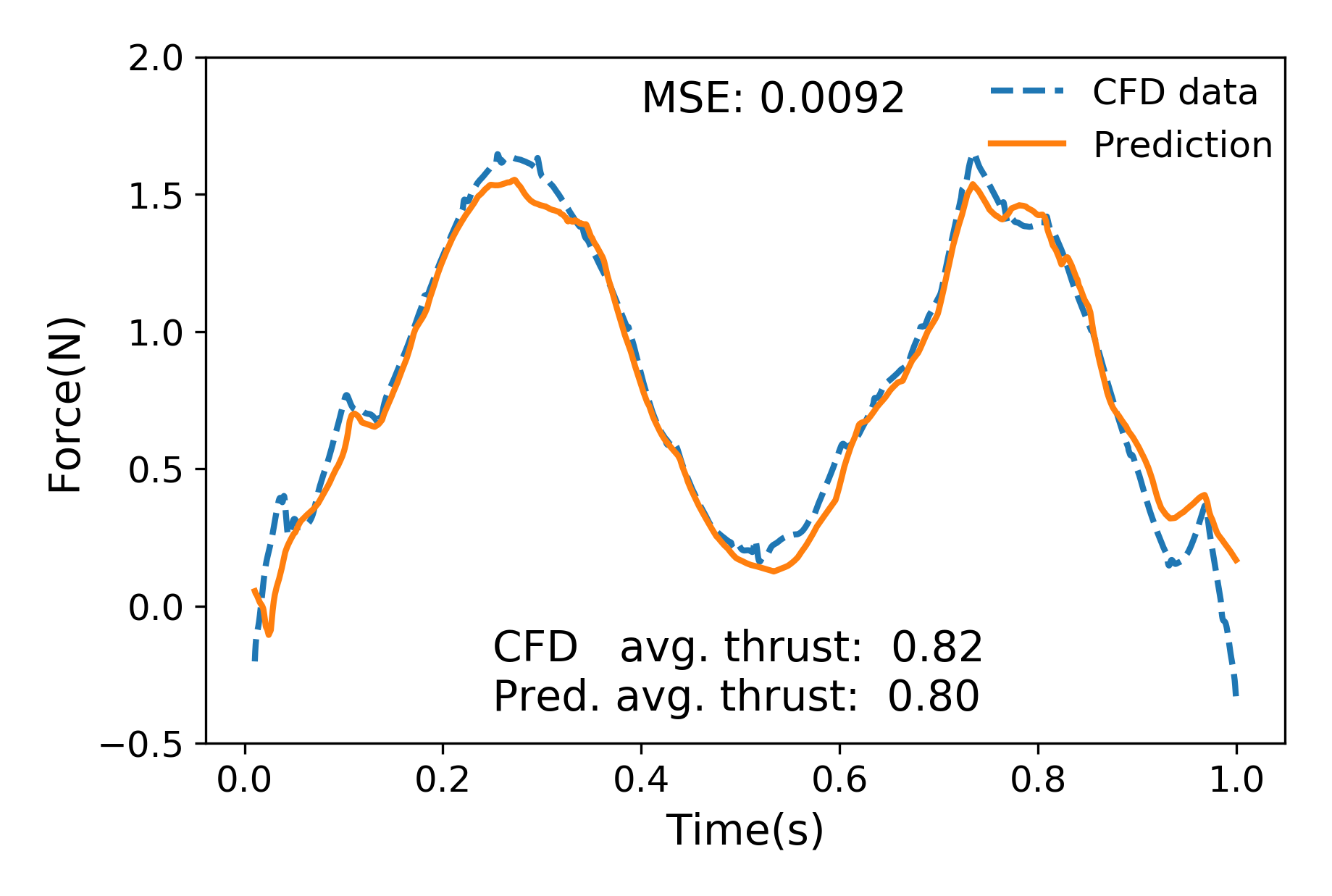}}}}
\caption{Comparison of two geometry-based models, plotted against CFD data.}
\label{fig-cnn1-pred2}
\end{figure}

To mitigate this error, our intuition was that the network needs more spatial data to better understand the relationship between the predicted force and the shape of the fin. To implement this, this new network is trained with ten chord lengths instead of five for both fin shapes. The hyperparameters are shown in table\ref{tab-caseE-hyp} and other parameters are the same as the previous experiment. Qualitatively, the predictions of the new model looks better than the one trained with just five. For validation, CFD simulations are done for both  bio-fin 1 and 2 shapes and fig.~\ref{fig-cnn1-pred2} shows the comparison with both network predictions. The impact of the geometric refinement is clearly visible with the 10-point geometry producing a better fit, fig.~\ref{fig-cnn2-bio2} compared with fig.~\ref{fig-cnn1-bio2-2}. Though predicted average thrust is lower(MSE gives a fit that is within 13\% of the peak to peak spread of 1.5 N), the ten-chord length model seems to exhibit a greater understanding of the relationship between a fin's shape and the force produced. This is further validated by the excellent fit(within 6\% of the peak to peak spread) of the bio-fin 1 prediction with the CFD result.

\begin{table}[htbp]
\caption{Case E: Hyperparameters}
\begin{center}
\begin{tabular}{|c|c|}
\hline
\textbf{Initial dense layer}& yes   \\
\hline
\textbf{Number of Layers}& 2   \\
\hline
\textbf{Number of units}& 32 \\
\hline
 \textbf{Activation Function}& \textit{Leaky ReLU}  \\
 \hline
\end{tabular}
\label{tab-caseE-hyp}
\end{center}
\end{table}

\subsection{Case F: Extension to the Rear Fin}

Finally, the above model is extended to a multi-fin configuration. This experiment uses two independently-trained models: a lead fin model that is identical to Case E, and a rear fin model that has additional inputs. The force production of the rear fin is more dependent on the fluid dynamics, as a result of the lead fin wake structures being incident on it, and thus requires more inputs as well. The spacing and phase offset between the two fins and the force output of the lead fin are added, since they directly impact the effect of the lead fin wake on the rear fin, and hence its force output. Table \ref{tab-caseF} shows the inputs for the rear fin network with column one inputs being specific to the single fin and column two inputs defining the single rear fin's dependence on the spatial configuration and output of the lead fin. The hyperparameters for the rear fin network are the same as the lead fin but without the initial dense layer.

\begin{table}[htbp]
\caption{Case F: Parameters}
\begin{center}
\begin{tabular}{|c|c|c|}
\hline
\multicolumn{2}{|c|}{\textbf{Input Parameters}}& \textbf{Output} \\
\cline{1-2}
\textbf{\textit{Rear fin}}& \textbf{\textit{+ data from Lead fin}}& \textbf{\textit{Predictions}} \\
\hline
10 geometric      & + X$_\text{offset}$ offset(spatial) from lead fin &  \\
chord lengths   & + stroke phase offset from lead fin &  \\
\cline{1-1}
Stroke and pitch  & & Force output  \\
angle time history & + Lead fin force output &\\
\cline{1-1}
Forward speed& &  \\
\hline
\multicolumn{3}{l}{$^{\mathrm{a}}$As shown in fig.~\ref{fig-geo-chord}.}
\end{tabular}
\label{tab-caseF}
\end{center}
\end{table}

\begin{figure}[htbp]
\centering \hspace{0in}
\mbox{\subfigure[10-point geometry comparison of bio-fin 1 rear]{\label{fig-cnn2-bio1-rear}{\includegraphics[width=2.8in,keepaspectratio]{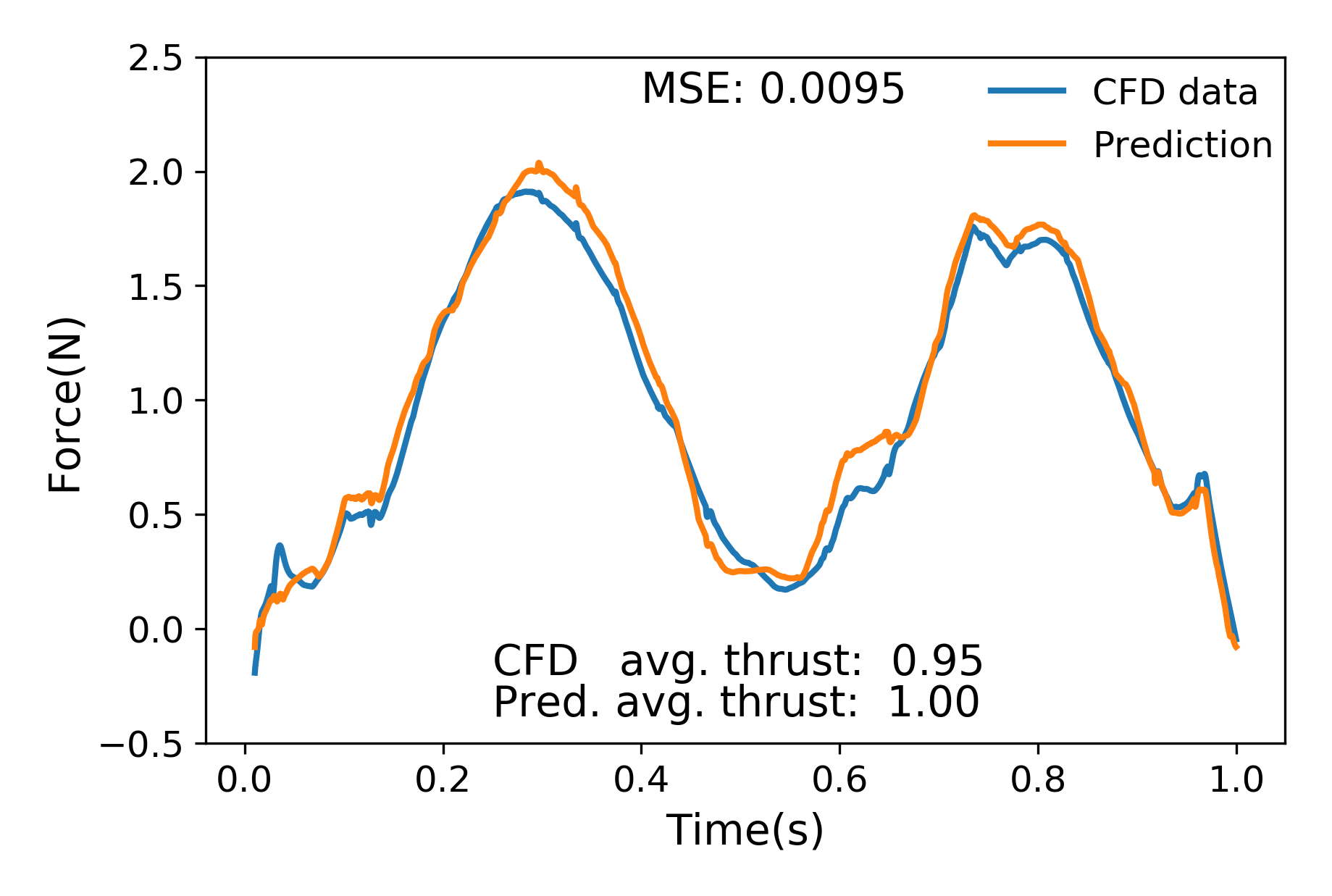}}}}
\hspace{0.in}
\mbox{\subfigure[10-point geometry comparison of bio-fin 2 rear]{\label{fig-cnn2-bio2-rear}{\includegraphics[width=2.8in,keepaspectratio]{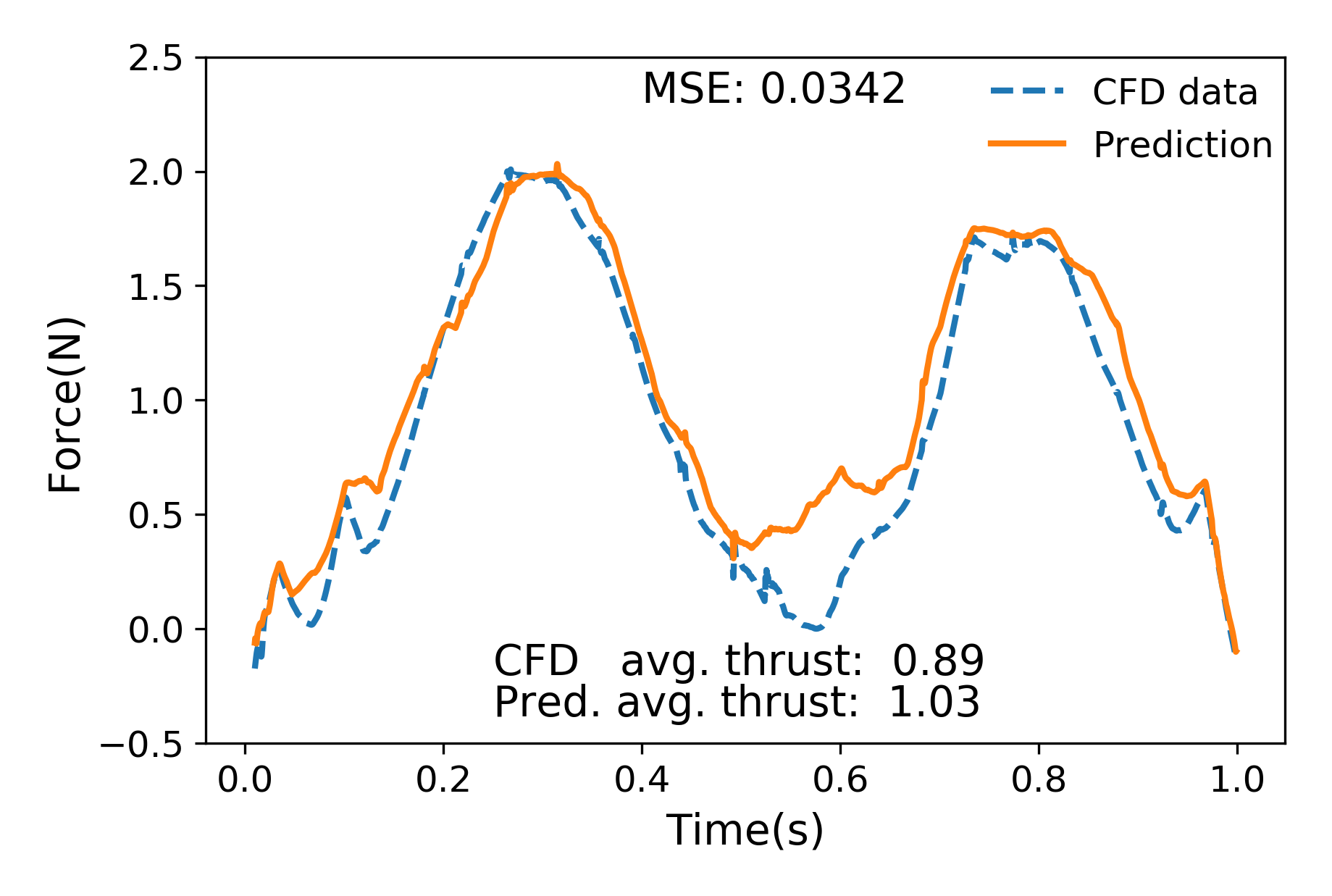}}}}
\caption{Comparison between prediction and CFD data for rear fin configurations.}
\label{fig-cnn2}
\end{figure}

After training, we again validate the predictions against simulation data. In figures~\ref{fig-cnn2-bio1-rear} and \ref{fig-cnn2-bio2-rear}, we see that qualitatively the predictions closely track the CFD data. For the bio-fin 2 run, we see a dip down to 0 N between the up/down strokes, that the network does not predict. Quantitatively, the network overpredicts the average rear fin thrust in both cases with the bio-fin 1 falling just within the rear fin experimental bounds of $\pm 0.05~\text{N}$  and bio-fin 2 rear having a larger deviation from the CFD results.

\subsection{Computational Performance}

As previously mentioned, the main objective of a surrogate model is to provide a fast approximation of some variables of interest given similar input framework as an experimental or CFD run. All cases tested showed a significant speedup over both experimental trials and CFD simulation.

The time required to collect data from experimental trials depends primarily on fin fabrication and measurement apparatus setup and calibration, that together takes on the order of days.  Once the experiment is set up, for our cases, collection of approximately 1000 stroke cycles takes about one hour. Post-processing the data to obtain forces takes a few more hours after that. A CFD simulation of a new geometry is significantly faster, but the process still incorporates grid generation, case setup, high performance computational resources, and post processing that still takes on the order of hours to days.
In contrast, predicting the force curve for a new fin configuration takes approximately 400 ms, that is four orders-of-magnitude faster than traditional data collection and five orders-of-magnitude faster than a full experimental trial. Training a neural network from scratch on our dataset (Cases C-F) takes several hours, depending on the specific architecture and input data format.

\section{Conclusion}

We demonstrate the development of neural networks, trained on existing experimental and CFD data, as surrogate reduced order models for fast accurate prediction of thrust forces generated through flapping fin propulsion. The results from multiple network architectures and different input frameworks indicate that neural networks are a viable option to handle the large input parameter space that span the kinematics, fin shapes, and fluid dynamic effects reflected in the output thrust forces.

The validation with CFD results show that the accuracy of the network predictions are within experimental bounds for the cycle averaged thrust and have low MSE's over a single cycle giving an accurate time history of the thrust profile for those cases that are closer to the training data in the parametric space. In the above cases, bio-fin 2 geometry has the biggest departure in shape and thus is the least accurate of the thrust profile model predictions. While still insightful, this accuracy needs to be further improved to enable a useful design and control tool. With this in mind, future steps will explore alternate loss functions during training and combined training for lead and rear fins. Further preprocessing of the training data and improving the input parameter space will also be investigated to improve the robustness of the trained networks. Given the outstanding performance gains with the above surrogate models, the tradeoff with accuracy, in our estimation favors these neural network models.

\section*{Acknowledgment}

This research has been sponsored by the Naval Research Laboratory (NRL) 6.2 Base Program. Computational and experimental resources were provided by the Laboratories for Computational Physics and Fluid Dynamics at NRL.

\bibliographystyle{IEEEtran}
\bibliography{IEEEabrv,ms}

\begin{thebibliography}{10}
\providecommand{\url}[1]{#1}
\csname url@samestyle\endcsname
\providecommand{\newblock}{\relax}
\providecommand{\bibinfo}[2]{#2}
\providecommand{\BIBentrySTDinterwordspacing}{\spaceskip=0pt\relax}
\providecommand{\BIBentryALTinterwordstretchfactor}{4}
\providecommand{\BIBentryALTinterwordspacing}{\spaceskip=\fontdimen2\font plus
\BIBentryALTinterwordstretchfactor\fontdimen3\font minus
  \fontdimen4\font\relax}
\providecommand{\BIBforeignlanguage}[2]{{%
\expandafter\ifx\csname l@#1\endcsname\relax
\typeout{** WARNING: IEEEtran.bst: No hyphenation pattern has been}%
\typeout{** loaded for the language `#1'. Using the pattern for}%
\typeout{** the default language instead.}%
\else
\language=\csname l@#1\endcsname
\fi
#2}}
\providecommand{\BIBdecl}{\relax}
\BIBdecl

\bibitem{hove_boxfishes_2001}
\BIBentryALTinterwordspacing
J.~Hove, L.~O'Bryan, M.~Gordon, P.~Webb, and D.~Weihs, ``Boxfishes
  ({Teleostei}: {Ostraciidae}) as a model system for fishes swimming with many
  fins: kinematics,'' \emph{Journal of Experimental Biology}, vol. 204, no.~8,
  p. 1459, Apr. 2001. [Online]. Available:
  \url{http://jeb.biologists.org/content/204/8/1459.abstract}
\BIBentrySTDinterwordspacing

\bibitem{lauder_learning_2006}
\BIBentryALTinterwordspacing
G.~V. Lauder and P.~G.~A. Madden, ``\BIBforeignlanguage{en}{Learning from fish:
  {Kinematics} and experimental hydrodynamics for roboticists},''
  \emph{\BIBforeignlanguage{en}{International Journal of Automation and
  Computing}}, vol.~3, no.~4, pp. 325--335, Oct. 2006. [Online]. Available:
  \url{http://link.springer.com/10.1007/s11633-006-0325-0}
\BIBentrySTDinterwordspacing

\bibitem{flammang_volumetric_2011}
\BIBentryALTinterwordspacing
B.~E. Flammang, G.~V. Lauder, D.~R. Troolin, and T.~E. Strand,
  ``\BIBforeignlanguage{en}{Volumetric imaging of fish locomotion},''
  \emph{\BIBforeignlanguage{en}{Biology Letters}}, vol.~7, no.~5, pp. 695--698,
  Oct. 2011. [Online]. Available:
  \url{https://royalsocietypublishing.org/doi/10.1098/rsbl.2011.0282}
\BIBentrySTDinterwordspacing

\bibitem{barrett_drag_1999}
D.~S. Barrett, M.~S. Triantafyllou, D.~K.~P. Yue, M.~A. Grosenbaugh, and M.~J.
  Wolfgang, ``\BIBforeignlanguage{en}{Drag reduction in fish-like
  locomotion},'' \emph{\BIBforeignlanguage{en}{Journal of Fluid Mechanics}},
  vol. 392, pp. 183--212, Aug. 1999.

\bibitem{licht_design_2004}
\BIBentryALTinterwordspacing
S.~Licht, V.~Polidoro, M.~Flores, F.~Hover, and M.~Triantafyllou,
  ``\BIBforeignlanguage{en}{Design and {Projected} {Performance} of a
  {Flapping} {Foil} {AUV}},'' \emph{\BIBforeignlanguage{en}{IEEE Journal of
  Oceanic Engineering}}, vol.~29, no.~3, pp. 786--794, Jul. 2004. [Online].
  Available: \url{http://ieeexplore.ieee.org/document/1353431/}
\BIBentrySTDinterwordspacing

\bibitem{kato_design_2008}
C.~Zhou, L.~Wang, Z.~Cao, S.~Wang, and M.~Tan, ``\BIBforeignlanguage{en}{Design
  and {Control} of {Biomimetic} {Robot} {Fish} {FAC}-{I}},'' in
  \emph{\BIBforeignlanguage{en}{Bio-mechanisms of {Swimming} and {Flying}}},
  N.~Kato and S.~Kamimura, Eds.\hskip 1em plus 0.5em minus 0.4em\relax Tokyo:
  Springer Japan, 2008, pp. 247--258.

\bibitem{sitorus_design_2009}
\BIBentryALTinterwordspacing
P.~E. Sitorus, Y.~Y. Nazaruddin, E.~Leksono, and A.~Budiyono,
  ``\BIBforeignlanguage{en}{Design and {Implementation} of {Paired} {Pectoral}
  {Fins} {Locomotion} of {Labriform} {Fish} {Applied} to a {Fish} {Robot}},''
  \emph{\BIBforeignlanguage{en}{Journal of Bionic Engineering}}, vol.~6, no.~1,
  pp. 37--45, Mar. 2009. [Online]. Available:
  \url{http://link.springer.com/10.1016/S1672-6529(08)60100-6}
\BIBentrySTDinterwordspacing

\bibitem{kato_elastic_2008}
N.~Kato, Y.~Ando, A.~Tomokazu, H.~Suzuki, K.~Suzumori, T.~Kanda, and S.~Endo,
  ``\BIBforeignlanguage{en}{Elastic {Pectoral} {Fin} {Actuators} for
  {Biomimetic} {Underwater} {Vehicles}},'' in
  \emph{\BIBforeignlanguage{en}{Bio-mechanisms of {Swimming} and {Flying}}},
  N.~Kato and S.~Kamimura, Eds.\hskip 1em plus 0.5em minus 0.4em\relax Tokyo:
  Springer Japan, 2008, pp. 271--282.

\bibitem{palmisano_robotic_2012}
J.~S. Palmisano, J.~D. Geder, R.~Ramamurti, W.~C. Sandberg, and R.~Banahalli,
  ``Robotic pectoral fin thrust vectoring using weighted gait combinations,''
  \emph{Applied Bionics and Biomechanics}, no.~3, pp. 333--345, 2012.

\bibitem{esposito_robotic_2012}
\BIBentryALTinterwordspacing
C.~J. Esposito, J.~L. Tangorra, B.~E. Flammang, and G.~V. Lauder,
  ``\BIBforeignlanguage{en}{A robotic fish caudal fin: effects of stiffness and
  motor program on locomotor performance},''
  \emph{\BIBforeignlanguage{en}{Journal of Experimental Biology}}, vol. 215,
  no.~1, pp. 56--67, Jan. 2012. [Online]. Available:
  \url{http://jeb.biologists.org/cgi/doi/10.1242/jeb.062711}
\BIBentrySTDinterwordspacing

\bibitem{moored_investigating_2008}
\BIBentryALTinterwordspacing
K.~W. Moored, W.~Smith, J.~Hester, W.~Chang, and H.~Bart-Smith, ``Investigating
  the {Thrust} {Production} of a {Myliobatoid}-{Inspired} {Oscillating}
  {Wing},'' \emph{Advances in Science and Technology}, vol.~58, pp. 25--30,
  Sep. 2008. [Online]. Available: \url{https://www.scientific.net/AST.58.25}
\BIBentrySTDinterwordspacing

\bibitem{bozkurttas_understanding_2008}
\BIBentryALTinterwordspacing
M.~Bozkurttas, J.~Tangorra, G.~Lauder, and R.~Mittal, ``Understanding the
  {Hydrodynamics} of {Swimming}: {From} {Fish} {Fins} to {Flexible}
  {Propulsors} for {Autonomous} {Underwater} {Vehicles},'' \emph{Advances in
  Science and Technology}, vol.~58, pp. 193--202, Sep. 2008. [Online].
  Available: \url{https://www.scientific.net/AST.58.193}
\BIBentrySTDinterwordspacing

\bibitem{akhtar_hydrodynamics_2007}
\BIBentryALTinterwordspacing
I.~Akhtar, R.~Mittal, G.~V. Lauder, and E.~Drucker,
  ``\BIBforeignlanguage{en}{Hydrodynamics of a biologically inspired tandem
  flapping foil configuration},'' \emph{\BIBforeignlanguage{en}{Theoretical and
  Computational Fluid Dynamics}}, vol.~21, no.~3, pp. 155--170, Apr. 2007.
  [Online]. Available: \url{http://link.springer.com/10.1007/s00162-007-0045-2}
\BIBentrySTDinterwordspacing

\bibitem{mignano_passing_2019}
\BIBentryALTinterwordspacing
A.~Mignano, S.~Kadapa, J.~Tangorra, and G.~Lauder,
  ``\BIBforeignlanguage{en}{Passing the {Wake}: {Using} {Multiple} {Fins} to
  {Shape} {Forces} for {Swimming}},''
  \emph{\BIBforeignlanguage{en}{Biomimetics}}, vol.~4, no.~1, p.~23, Mar. 2019.
  [Online]. Available: \url{https://www.mdpi.com/2313-7673/4/1/23}
\BIBentrySTDinterwordspacing

\bibitem{ramamurti_computational_2018}
\BIBentryALTinterwordspacing
R.~Ramamurti, J.~Geder, K.~Viswanath, and M.~Pruessner,
  ``\BIBforeignlanguage{en}{Computational {Fluid} {Dynamics} {Study} of the
  {Propulsion} {Characteristics} of {Tandem} {Flapping} {Fins}},'' in
  \emph{\BIBforeignlanguage{en}{2018 {AIAA} {Aerospace} {Sciences}
  {Meeting}}}.\hskip 1em plus 0.5em minus 0.4em\relax Kissimmee, Florida:
  American Institute of Aeronautics and Astronautics, Jan. 2018. [Online].
  Available: \url{https://arc.aiaa.org/doi/10.2514/6.2018-0039}
\BIBentrySTDinterwordspacing

\bibitem{geder_underwater_2017}
\BIBentryALTinterwordspacing
J.~D. Geder, R.~Ramamurti, K.~Viswanath, and M.~Pruessner,
  ``\BIBforeignlanguage{English}{Underwater thrust performance of tandem
  flapping fins: {Effects} of stroke phasing and fin spacing},'' in
  \emph{\BIBforeignlanguage{English}{{MTS}/{IEEE} {OCEANS} '17
  {Anchorage}}}.\hskip 1em plus 0.5em minus 0.4em\relax Anchorage, Alaska:
  Institute of Electrical and Electronics Engineers, Sep. 2017, oCLC:
  1020270128. [Online]. Available:
  \url{http://ieeexplore.ieee.org/xpl/mostRecentIssue.jsp?punumber=8190426}
\BIBentrySTDinterwordspacing

\bibitem{ramamurti_propulsion_2019}
\BIBentryALTinterwordspacing
R.~Ramamurti, J.~Geder, K.~Viswanath, and M.~Pruessner,
  ``\BIBforeignlanguage{en}{Propulsion {Characteristics} of {Flapping} {Caudal}
  {Fins} and its {Upstream} {Interaction} with {Pectoral} {Fins}},'' in
  \emph{\BIBforeignlanguage{en}{{AIAA} {Scitech} 2019 {Forum}}}.\hskip 1em plus
  0.5em minus 0.4em\relax San Diego, California: American Institute of
  Aeronautics and Astronautics, Jan. 2019. [Online]. Available:
  \url{https://arc.aiaa.org/doi/10.2514/6.2019-1618}
\BIBentrySTDinterwordspacing

\bibitem{geder_four-fin_2011}
J.~D. Geder, R.~Ramamurti, J.~Palmisano, M.~Pruessner, B.~Ratna, and W.~C.
  Sandberg, ``Four-{Fin} {Bio}-{Inspired} {UUV}: {Modeling} and {Control}
  {Solutions},'' in \emph{Volume 2: {Biomedical} and {Biotechnology}
  {Engineering}; {Nanoengineering} for {Medicine} and {Biology}}.\hskip 1em
  plus 0.5em minus 0.4em\relax Denver, Colorado, USA: ASMEDC, Jan. 2011, pp.
  799--808.

\bibitem{muscutt_performance_2017}
L.~E. Muscutt, G.~D. Weymouth, and B.~Ganapathisubramani,
  ``\BIBforeignlanguage{en}{Performance augmentation mechanism of in-line
  tandem flapping foils},'' \emph{\BIBforeignlanguage{en}{Journal of Fluid
  Mechanics}}, vol. 827, pp. 484--505, Sep. 2017.

\bibitem{liebe_diversity_2006}
\BIBentryALTinterwordspacing
F.~Fish, ``\BIBforeignlanguage{en}{Diversity, mechanics and performance of
  natural aquatic propulsors},'' in \emph{\BIBforeignlanguage{en}{{WIT}
  {Transactions} on {State} of the {Art} in {Science} and {Engineering}}},
  1st~ed., R.~Liebe, Ed.\hskip 1em plus 0.5em minus 0.4em\relax WIT Press, Nov.
  2006, vol.~1, pp. 57--87. [Online]. Available:
  \url{http://library.witpress.com/viewpaper.asp?pcode=1845640012-201-1}
\BIBentrySTDinterwordspacing

\bibitem{Ged18}
J.~D. Geder, R.~Ramamurti, K.~Viswanath, M.~Pruessner, and R.~Koehler, ``Effect
  of {{Flow Interaction}} between {{Median Paired}} and {{Caudal Fins}} on
  {{Propulsion}},'' in \emph{{{MTS}}/{{IEEE OCEANS Conference}} '18},
  {Charleston, SC}, Oct. 2018.

\bibitem{Hor89}
K.~Hornik, M.~Stinchcombe, and H.~White, ``\BIBforeignlanguage{en}{Multilayer
  feedforward networks are universal approximators},''
  \emph{\BIBforeignlanguage{en}{Neural Networks}}, vol.~2, no.~5, pp. 359--366,
  Jan. 1989.

\bibitem{Kri12}
A.~Krizhevsky, I.~Sutskever, and G.~E. Hinton, ``Imagenet classification with
  deep convolutional neural networks,'' in \emph{Advances in Neural Information
  Processing Systems}, 2012, pp. 1097--1105.

\bibitem{Mik13a}
T.~Mikolov, I.~Sutskever, K.~Chen, G.~S. Corrado, and J.~Dean, ``Distributed
  representations of words and phrases and their compositionality,'' in
  \emph{Advances in Neural Information Processing Systems}, 2013, pp.
  3111--3119.

\bibitem{Rai19}
M.~Raissi, P.~Perdikaris, and G.~E. Karniadakis,
  ``\BIBforeignlanguage{en}{Physics-informed neural networks: {{A}} deep
  learning framework for solving forward and inverse problems involving
  nonlinear partial differential equations},''
  \emph{\BIBforeignlanguage{en}{Journal of Computational Physics}}, vol. 378,
  pp. 686--707, Feb. 2019.

\bibitem{Tri18a}
R.~Tripathy and I.~Bilionis, ``\BIBforeignlanguage{en}{Deep {{UQ}}:
  {{Learning}} deep neural network surrogate models for high dimensional
  uncertainty quantification},'' \emph{\BIBforeignlanguage{en}{Journal of
  Computational Physics}}, vol. 375, pp. 565--588, Dec. 2018.

\bibitem{Zhu18a}
Y.~Zhu and N.~Zabaras, ``\BIBforeignlanguage{en}{Bayesian {{Deep Convolutional
  Encoder}}-{{Decoder Networks}} for {{Surrogate Modeling}} and {{Uncertainty
  Quantification}}},'' \emph{\BIBforeignlanguage{en}{Journal of Computational
  Physics}}, vol. 366, pp. 415--447, Aug. 2018.

\bibitem{Maa13}
A.~L. Maas, A.~Y. Hannun, and A.~Y. Ng, ``\BIBforeignlanguage{en}{Rectifier
  {{Nonlinearities Improve Neural Network Acoustic Models}}},'' in
  \emph{\BIBforeignlanguage{en}{International {{Conference}} on {{Machine
  Learning}}}}, 2013, p.~6.

\bibitem{Sri14}
N.~Srivastava, G.~Hinton, A.~Krizhevsky, I.~Sutskever, and R.~Salakhutdinov,
  ``\BIBforeignlanguage{en}{Dropout: {{A Simple Way}} to {{Prevent Neural
  Networks}} from {{Overfitting}}},'' \emph{\BIBforeignlanguage{en}{Journal of
  Machine Learning Research}}, p.~30, 2014.

\bibitem{Kin15a}
D.~P. Kingma and J.~Ba, ``Adam: {{A}} method for stochastic optimization,'' in
  \emph{International {{Conference}} on {{Learning Representations}}}, 2015.

\bibitem{tensorflow}
M.~Abadi, A.~Agarwal, P.~Barham, E.~Brevdo, Z.~Chen, C.~Citro, G.~S. Corrado,
  A.~Davis, J.~Dean, M.~Devin, S.~Ghemawat, I.~Goodfellow, A.~Harp, G.~Irving,
  M.~Isard, {Yangqing Jia}, R.~Jozefowicz, L.~Kaiser, M.~Kudlur, J.~Levenberg,
  D.~Man{\'e}, R.~Monga, S.~Moore, D.~Murray, C.~Olah, M.~Schuster, J.~Shlens,
  B.~Steiner, I.~Sutskever, K.~Talwar, P.~Tucker, V.~Vanhoucke, V.~Vasudevan,
  F.~Vi{\'e}gas, O.~Vinyals, P.~Warden, M.~Wattenberg, M.~Wicke, Y.~Yu, and
  X.~Zheng, ``{{TensorFlow}}: {{Large}}-scale machine learning on heterogeneous
  systems,'' 2015.

\end{thebibliography}

\end{document}